\documentclass[10pt]{article}
\usepackage{arxiv}
\usepackage[T1]{fontenc}
\usepackage[utf8]{inputenc}
\usepackage{amsmath,amssymb,amsfonts,amsthm}
\usepackage{algorithm}
\usepackage{algpseudocode}
\usepackage{graphicx}
\usepackage{booktabs}
\usepackage{multirow}
\usepackage{array}
\usepackage{cite}
\usepackage{url}
\usepackage{xcolor}
\usepackage{listings}
\usepackage[colorlinks=true,linkcolor=blue!65!black,citecolor=green!45!black,urlcolor=blue!70!black]{hyperref}

\newtheorem{lemma}{Lemma}
\newtheorem{proposition}{Proposition}
\newtheorem{observation}{Observation}

\newcommand{\sysname}{\textsc{TraceLift}}

\begin{document}

\title{\sysname{}: Execution-Distilled Binary Deobfuscation via Temporal Waypoint Repair and Two-Link Verification}

\author{
  \textbf{Junsheng Fu} \qquad \textbf{Shuangshuang Gao} \\
  Zhengzhou University of Industrial Technology \\
  Zhengzhou, China \\
  \texttt{23030130144@stu.zzuit.edu.cn}
}

\maketitle

\begin{abstract}
Composite binary obfuscation combines Control-Flow Flattening (FLA), Bogus Control Flow (BCF), Mixed Boolean-Arithmetic (MBA), and runtime Self-Modifying Code (SMC). Together, these passes severely degrade both heuristic and Large Language Model (LLM) decompilers. Although classical dynamic slicers and algebraic simplifiers handle individual passes at the instruction level, lifting execution traces into high-level C or Python loops fails across two interface bottlenecks. First, self-erasing packers overwrite or zero out code bytes immediately after execution ($a@v_1 \to a@v_2 \to \texttt{0x00}$), while algebraic taint laundering wires zero-valued invariants into output registers to fool syntactic backward slicing. Second, when an LLM folds an execution trace into a loop, a single-step arithmetic error triggers multi-round bitwise avalanche. End-to-end I/O counterexamples then exhibit scrambled residuals ($\approx w/2$ Hamming distance) that hide the faulty round.

We present \sysname{}, a training-free neuro-symbolic deobfuscation framework that resolves both bottlenecks. During extraction, \sysname{} records per-fetch code snapshots before memory erasure and prunes control-scaffolding and taint-laundering subgraphs prior to \textsc{SiMBA} algebraic canonicalization. During repair, \sysname{} introduces \emph{Temporal Waypoint Alignment} to compare binary loop back-edge snapshots against candidate loop states and isolate the earliest divergent round $r^*$ before avalanche diffusion. Candidate reconstructions are certified by a two-link verification chain. This chain pairs trace-projected Z3 equivalence over fixed-length buffers with a relative AST branch guard and 100 per-function unseen fuzzing inputs. Across 48 compiled GCC (\texttt{-O0} to \texttt{-O3}) binaries spanning 12 cryptographic and hash families, \sysname{} prunes $59.8\%$ of dynamic obfuscation instructions and reduces CFG cyclomatic complexity by $60.0\%$--$87.5\%$ (mean $76.8\%$). On a live local 7B coder model (\texttt{Qwen2.5-Coder-7B}), \sysname{} raises functional equivalence from $0.0\%$ ($0/12$ under static one-shot and a \textsc{SymDec}-style feedback baseline) to $91.7\%$ ($11/12$) within 3 repair turns.
\end{abstract}


\section{Introduction}
Automated binary deobfuscation is a core requirement in malware forensics, software protection auditing, and vulnerability discovery~\cite{shoshitaishvili2016sok, yadegari2015generic, hu2026bindeobfbench}. In practice, software protectors and malware authors rarely apply a single transformation in isolation. Instead, they compose compile-time structural passes, such as Obfuscator-LLVM (OLLVM) Control-Flow Flattening (FLA) and Bogus Control Flow (BCF)~\cite{junod2015ollvm, ming2015loop}, with data-flow passes like Mixed Boolean-Arithmetic (MBA)~\cite{rakotonirina2022simba, blazytko2017syntia} and runtime Self-Modifying Code (SMC)~\cite{coogan2011deobfuscation, ugarte2015deep}.

Recent neural decompilation systems use code-pretrained LLMs to translate low-level assembly or decompiler pseudocode into readable C functions~\cite{tan2024llm4decompile, hu2024degpt, hu2026symdec, li2026bindec, jiang2026lohre}. These models perform well on clean compiler outputs (\texttt{O0}--\texttt{O3}). However, recent empirical evaluations (\textsc{BinDeObfBench}~\cite{hu2026bindeobfbench}, M{\"a}chtle \emph{et al.}~\cite{maechtle2026rethinking}, and Tkachenko \emph{et al.}~\cite{tkachenko2025deconstructing}) show that LLM accuracy collapses under composite obfuscation. Single-pass functional equivalence drops below $17\%$ on individual passes and reaches $0\%$ when FLA, BCF, MBA, and SMC are combined.

\begin{figure}[!htbp]
\centering
\includegraphics[width=0.98\textwidth]{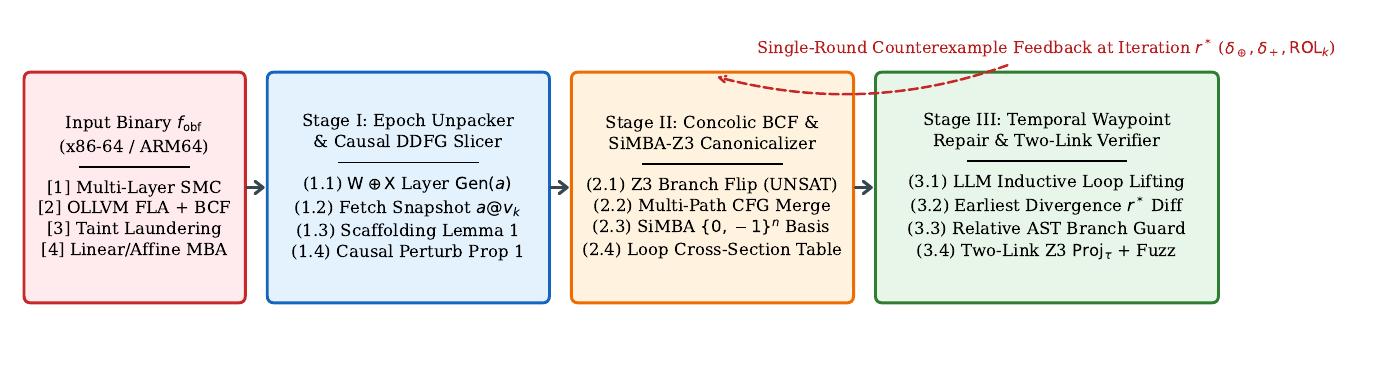}
\caption{Overview of \sysname{}. Stage~I captures per-fetch epoch-versioned $\text{W}\oplus\text{X}$ snapshots ($a@v_k$) and prunes closed control-scaffolding and taint-laundering subgraphs. Stage~II applies Concolic BCF branch flipping and \textsc{SiMBA} $\{0,-1\}^n$ linear MBA decomposition. Stage~III guides inductive loop synthesis via Temporal Waypoint Alignment ($r^*$) and a two-link verification chain ($\text{Proj}_\tau$ Z3 proof + held-out Unicorn fuzzing).}
\label{fig:arch}
\end{figure}

\subsection{The Neuro-Symbolic Interface Gap}
Classical program analysis has produced strong primitives for individual obfuscation layers. These include dynamic Def-Use backward slicing~\cite{coogan2011deobfuscation, yadegari2015generic}, SMT-based opaque predicate elimination (\textsc{LOOP}~\cite{ming2015loop}), linear MBA basis decomposition (\textsc{SiMBA}~\cite{rakotonirina2022simba}), and dynamic LLVM lifting (\textsc{BinRec}~\cite{altinay2020binrec}, \textsc{SATURN}~\cite{garba2019saturn}). Why, then, does recovering verifiable high-level source code from composite obfuscated binaries remain difficult?

Our key observation is that \emph{the primary bottleneck lies at the interface between low-level dynamic execution and high-level neural loop induction}, rather than in isolated deobfuscation primitives. Two structural mismatches occur at this boundary.

First, \emph{inductive loop lifting is defeated by multi-round avalanche}. Classical dynamic lifters such as \textsc{BinRec}~\cite{altinay2020binrec} emit thousands of lines of unrolled LLVM IR rather than compact source loops. When an LLM inductively folds a trace into a \texttt{for} or \texttt{while} loop, a small arithmetic error in Round~0 (such as a wrong shift mask or magic-multiplier constant) diffuses rapidly across subsequent rounds~\cite{webster1985design}. Existing symbolic repair systems such as \textsc{SymDec}~\cite{hu2026symdec} compare only end-to-end function outputs. After four or more mixing rounds, the output residual exhibits a near-random Hamming distance ($\approx w/2$ bits). An LLM cannot invert this scrambled residual to locate the single-step fault. Meanwhile, running whole-binary symbolic execution directly on a flattened binary triggers path explosion inside the dispatcher loop~\cite{banescu2016code}.

Second, \emph{post-mortem unpacking and syntactic slicing leak adversarial noise into the LLM context}. Traditional unpackers dump memory after reaching a single Original Entry Point (OEP)~\cite{ugarte2015deep}. However, wave-by-wave self-erasing SMC executes decrypted bytes at address $a$, overwrites $a$ in the next wave ($a@v_1 \to a@v_2$), and zeroes out $a$ (\texttt{0x00}) before returning. Post-mortem dumps miss earlier waves entirely. In addition, attackers can defeat syntactic Def-Use slicing~\cite{yadegari2015generic} via \emph{algebraic taint laundering}. By computing a multi-instruction opaque zero invariant $r_{10} = (((x+1)x) \land 1) \times \text{0x1337} \equiv 0$ and adding it to the output register ($y \leftarrow y + r_{10}$), the attacker creates a genuine syntactic Def-Use edge into the output sink. Syntactic backward slicing retains the entire parasitic subgraph and overwhelms small and medium code LLMs.

\subsection{Our Contributions}
\sysname{} (Fig.~\ref{fig:arch}) bridges established dynamic analysis primitives~\cite{yadegari2015generic, ming2015loop, rakotonirina2022simba} with LLM loop synthesis by resolving both interface mismatches:
\begin{itemize}
    \item \textbf{Temporal Waypoint Repair and Two-Link Verification (Section~\ref{sec:method_verify}):} We formalize multi-round bit-diffusion masking (Observation~\ref{obs:avalanche}) and align binary loop latch snapshots against candidate loop states to pinpoint the earliest divergent iteration $r^*$. We also formulate an explicit two-link verification chain ($f_{\text{obf}} \to f_{\text{clean}} \to \hat{f}$) that combines trace-projected Z3 equivalence over fixed-length buffers with a relative AST branch guard and held-out Unicorn fuzzing.
    \item \textbf{Per-Fetch Epoch-Versioned Unpacking and Causal Slicing (Section~\ref{sec:method_slice}):} We design a full-address-space $\text{W}\oplus\text{X}$ unpacker that records per-fetch snapshots $(a, v_{\text{epoch}}, \text{Gen}(a))$ to recover overwritten and self-erased instruction waves. We establish a threshold-free closed control-scaffolding pruning criterion (Lemma~\ref{lem:scaffolding}), characterize its completeness boundary under pointer aliasing, and use causal value perturbation (Proposition~\ref{prop:causal}) to excise zero-entropy taint-laundering cones.
    \item \textbf{Prior-Art Grounded Algebraic Canonicalization (Section~\ref{sec:method_mba}):} We integrate trace-guided Concolic branch flipping~\cite{ming2015loop} with \textsc{SiMBA}'s $\{0,-1\}^n$ Boolean-lattice linear MBA decomposition~\cite{rakotonirina2022simba}, annotating distilled traces with Z3-verified algebraic expressions and Loop Cross-Section State Tables.
    \item \textbf{Evaluation Against a \textsc{SymDec}-Style Baseline and Archived CTF Binaries (Section~\ref{sec:eval}):} We evaluate \sysname{} on 48 compiled GCC (\texttt{-O0} to \texttt{-O3}) binaries across 12 algorithm families, \texttt{ARM64} code, live local Ollama LLMs (\texttt{Qwen2.5-Coder-7B} and \texttt{Qwen3-8B}), and four unmodified Linux CTF ELF binaries. \sysname{} improves local 7B functional recovery from $0.0\%$ ($0/12$ under a \textsc{SymDec}-style end-to-end feedback baseline) to $91.7\%$ ($11/12$).
\end{itemize}

\section{Background and Motivating Example}
\label{sec:motivation}
Consider a representative stream-hash function $f_{\text{obf}}(\texttt{buf}, \texttt{len})$ protected by four composed obfuscation layers. Layer~1 is an outer runtime SMC XOR-unpacker stub (\texttt{0x400000..0x40003f}). Layer~2 is an OLLVM control-flow flattening dispatcher controlled by a stack state variable \texttt{[rbp-0x14]}. Layer~3 injects a number-theoretic bogus branch $((x(x+1)) \land 1) \neq 0$ and launders the dead variable \texttt{r10} into the output stream. Layer~4 obfuscates the byte accumulator using a linear MBA identity and an 8-bit left rotation ($\text{ROL}_3$).

\begin{figure}[!htbp]
\centering
\begin{minipage}[t]{0.32\textwidth}
\begin{lstlisting}[title={(a) Baseline Static Decompiler / Raw ASM}]
// 0x400000: Outer SMC Unpacker Stub
void f_obf(uint8_t *rdi, uint64_t rsi) {
  uint8_t *p = (uint8_t *)0x400040;
  for (int k = 0; k < 144; k++)
    p[k] ^= 0xA7; // Modifies code in-place
  // Static IDA/Ghidra on disk sees encrypted
  // bytes at 0x400040: db 0x1e, 0x8f, 0x99...
  // Even if manually unpacked, static F5 emits:
  uint32_t state = 0x1001;
  while (1) {
    if (state == 0x1001) {
      if (rcx >= rsi) state = 0x4004;
      else state = 0x2002;
    } else if (state == 0x2002) {
      uint32_t x = rdi[rcx];
      // Opaque BCF + Taint Laundering:
      uint32_t r10 = ((x + 1) * x) & 1;
      if (r10 != 0) { rax ^= 0x1337BEEF; break; }
      r10 *= 0x1337; // Syntactic taint edge
      edx = ((x ^ 0x5A) ^ rcx) + 2*((x ^ 0x5A) & rcx) + r10;
      state = 0x3003;
    } else if (state == 0x3003) { ... }
  }
}
// -> One-shot Static LLM Output:
// Fails to unpack 0x400040 (0% pass rate);
// retains dead r10 branch and omits ROL-3.
\end{lstlisting}
\end{minipage}\hfill
\begin{minipage}[t]{0.33\textwidth}
\begin{lstlisting}[title={(b) \sysname{} Distilled Context (50.2$\times$)}]
=== TRACELIFT DISTILLED CONTEXT ===
[Unpacker] L0 Stub (0x400000..0x40003f) stripped.
[DDFG+Lem1] Pruned 6 FLA Dispatchers & [rbp-0x14].
[CausalPerturb] Excised r10 taint-launder cone
  (0x40000d..0x400025): Delta_Out == 0 (H = 0).
[Concolic-Z3] Branch 0x40008e: UNSAT (Opaque BCF).
[SiMBA-Z3] (u^v)+2*(u&v) => u + v [PROVEN UNSAT]

--- Core Payload Blocks (17 Insns) ---
Block_0x400040 (Entry):
  xor eax, eax          ; acc = 0
  xor ecx, ecx          ; i = 0
Block_0x400085 (Loop Body, hits=8):
  movzx edx, byte [rdi + rcx]
  ; [SiMBA-Z3 Certificate]: edx = (rdi[rcx]^0x5a)+rcx
  xor edx, 0x5a
  add edx, ecx
  and edx, 0xff
  ; 8-bit Rotate Left by 3 (shl 3 | shr 5):
  mov ebx, edx; shl ebx, 3; shr edx, 5; or ebx, edx
  and ebx, 0xff
  mov byte [rdi + rcx], bl
  add rax, rbx; inc rcx

--- Loop Cross-Section Table (Input=00..07) ---
Iter r=0: in=0x00 -> pre_rol=0x5a -> bl=0xd2, acc=0x0d2
Iter r=1: in=0x01 -> pre_rol=0x5c -> bl=0xe2, acc=0x1b4
Iter r=2: in=0x02 -> pre_rol=0x5a -> bl=0xd2, acc=0x286
\end{lstlisting}
\end{minipage}\hfill
\begin{minipage}[t]{0.32\textwidth}
\begin{lstlisting}[title={(c) Stage III: Waypoint-Guided Repair}]
// [Turn 1 Candidate by Local 7B LLM]:
// Lifts XOR and index add, but omits ROL-3:
uint64_t cand_turn1(uint8_t *buf, uint64_t len) {
  uint64_t acc = 0;
  for (uint64_t i = 0; i < len; i++) {
    uint8_t v = ((buf[i] ^ 0x5A) + i) & 0xFF;
    buf[i] = v; acc += v;
  }
  return acc;
}
// [TRACELIFT Temporal Waypoint Diagnosis]:
// Earliest Divergence Pinpointed at Iter r* = 0.
// Pre-State: AGREED. At r* = 0 (in[0] = 0x00):
//   Expected buf[0] = 0xd2, Got = 0x5a
//   (XOR_diff = 0x88).
//   Bit-Alignment Hint: ROL8(0x5a, 3) == 0xd2.

// [Turn 2 Repaired Candidate]:
uint64_t cand_turn2(uint8_t *buf, uint64_t len) {
  uint64_t acc = 0;
  for (uint64_t i = 0; i < len; i++) {
    uint8_t v = ((buf[i] ^ 0x5A) + i) & 0xFF;
    v = (uint8_t)((v << 3) | (v >> 5));
    buf[i] = v; acc += v;
  }
  return acc;
}
// [Verification]: Probes PASS | AST Branch Guard PASS
// Fuzz: 100/100 (1000/1000 Stress) | Z3 Proj: UNSAT
\end{lstlisting}
\end{minipage}
\caption{Motivating comparison on a composite SMC + FLA + BCF + Taint-Laundering + MBA function: (a) Static decompilation and one-shot LLM failure; (b) \sysname{}'s distilled context ($854 \to 17$ instructions, $50.2\times$ compression) with \textsc{SiMBA}-Z3 annotations and Loop Cross-Section Table; (c) Temporal Waypoint fault localization at iteration $r^*=0$ guiding repair.}
\label{fig:motivating_code}
\end{figure}

As shown in Fig.~\ref{fig:motivating_code}(a), static decompilers and one-shot LLMs fail at every layer. On disk, the payload at \texttt{0x400040} contains encrypted bytes. Even if unpacked manually, the flattened dispatcher and laundered taint variable \texttt{r10} cause the LLM to retain dead branches and omit the 8-bit rotation.

Fig.~\ref{fig:motivating_code}(b) shows how \sysname{} distills this execution. Stage~I strips the L0 unpacker stub, prunes the six FLA dispatcher blocks via Lemma~\ref{lem:scaffolding}, and excises the zero-entropy \texttt{r10} cone via Proposition~\ref{prop:causal}. Stage~II proves the BCF branch \texttt{UNSAT} in Z3, simplifies the MBA expression into $u+v$ via \textsc{SiMBA}, and records the Loop Cross-Section Table. When the Turn-1 candidate omits the 3-bit rotation (Fig.~\ref{fig:motivating_code}(c)), Stage~III isolates the divergence at iteration $r^*=0$ and reports the exact bit-rotation residual ($\text{ROL}_8(\texttt{0x5a}, 3) = \texttt{0xd2}$). The repaired Turn-2 candidate passes the relative AST branch guard, 100 per-function held-out fuzz inputs (plus 1{,}000 stress-test vectors), and trace-projected Z3 verification.

\section{Related Work}
Table~\ref{tab:related_work} summarizes representative classical deobfuscators and recent 2024--2026 neural decompilation frameworks, highlighting their core mechanisms and the specific interface gap each leaves open under composite binary obfuscation.

\begin{table}[!htbp]
\centering
\caption{Comparison of Representative Classical Deobfuscators and Neural Decompilation Frameworks}
\label{tab:related_work}
\footnotesize
\setlength{\tabcolsep}{4.5pt}
\resizebox{\textwidth}{!}{%
\begin{tabular}{ll>{\raggedright\arraybackslash}p{5.2cm}>{\raggedright\arraybackslash}p{5.8cm}}
\toprule
\textbf{System / Study} & \textbf{Venue} & \textbf{Primary Technical Mechanism} & \textbf{Interface Limitation Under Composite Obfuscation} \\
\midrule
Coogan \emph{et al.}~\cite{coogan2011deobfuscation} & CCS'11 & Equational dynamic trace slicing for virtualization obfuscators & Emits low-level instruction slices without structured source loop recovery \\
Yadegari \emph{et al.}~\cite{yadegari2015generic} & S\&P'15 & Bit-level taint propagation and syntactic backward Def-Use slicing & Retains zero-entropy algebraic taint-laundering cones wired into output sinks \\
\textsc{LOOP}~\cite{ming2015loop} & CCS'15 & Path-constrained SMT solving for logic-oriented opaque predicates & Targets BCF predicates only; does not strip FLA dispatchers or multi-wave SMC \\
\textsc{Syntia}~\cite{blazytko2017syntia} / \textsc{QSynth}~\cite{david2020qsynthesis} & USENIX'17 & Monte Carlo Tree Search and lookup-table synthesis for MBA expressions & Operates on isolated expressions; stochastic search degrades on nested affine MBAs \\
\textsc{SATURN}~\cite{garba2019saturn} / \textsc{BinRec}~\cite{altinay2020binrec} & EuroSys'20 & Dynamic binary lifting to LLVM IR followed by \texttt{-O3} optimization passes & Produces unrolled LLVM IR traces rather than human-readable high-level C loops \\
\textsc{SiMBA}~\cite{rakotonirina2022simba} & CCS'22 & Exact $\{0,-1\}^n$ Boolean-lattice linear MBA basis decomposition & Simplifies stateless bitwise-arithmetic expressions; does not address control-flow flattening \\
\textsc{DeGPT}~\cite{hu2024degpt} / \textsc{LLM4Decompile}~\cite{tan2024llm4decompile} & NDSS'24 & One-shot static LLM translation and decompiler pseudocode refinement & Static-only input; accuracy drops by $>70\%$ under OLLVM FLA/BCF~\cite{maechtle2026rethinking} \\
Tkachenko \emph{et al.}~\cite{tkachenko2025deconstructing} / \textsc{BinDeObfBench}~\cite{hu2026bindeobfbench} & arXiv'25--26 & Empirical benchmarks and supervised fine-tuning on obfuscated assembly/pseudocode & Static translation without dynamic unpacking, causal slicing, or execution repair \\
M{\"a}chtle \emph{et al.}~\cite{maechtle2026rethinking} & ICICS'26 & Multi-sample \texttt{Pass@100} candidate generation and execution ranking & Assumes pre-existing unit tests; fixed test suites remain vulnerable to lookup overfitting \\
\textsc{LOHRE}~\cite{jiang2026lohre} & arXiv'26 & Execution-grounded user-defined type (UDT) layout recovery via depth heuristics & Recovers struct memory layouts rather than deobfuscating flattened or packed control flow \\
\textsc{BinDec}~\cite{li2026bindec} / \textsc{SymDec}~\cite{hu2026symdec} & JoS / arXiv'26 & Coverage-guided symbolic execution and end-to-end I/O counterexample feedback & Evaluated on clean \texttt{O0--O3}; whole-binary SMT explodes on FLA and end-to-end diffs fail under round avalanche \\
\midrule
\textbf{\sysname{} (This Work)} & \textbf{2026} & \textbf{Per-fetch epoch unpacking ($a@v_k$), causal DDFG slicing, \textsc{SiMBA} basis canonicalization, and Temporal Waypoint ($r^*$) repair} & \textbf{Bridges dynamic trace distillation with single-round fault localization and two-link ($\text{Proj}_\tau$ Z3 + fuzz) verification} \\
\bottomrule
\end{tabular}%
}
\end{table}

\subsection{Classical Dynamic Deobfuscation and Algebraic Simplification}
Coogan \emph{et al.}~\cite{coogan2011deobfuscation} and Yadegari \emph{et al.}~\cite{yadegari2015generic} established dynamic backward slicing and bit-level taint propagation to strip virtualization and flattening dispatchers. Ming \emph{et al.} (\textsc{LOOP}~\cite{ming2015loop}) used path-constrained SMT queries to detect opaque predicates. For data-flow obfuscation, \textsc{Syntia}~\cite{blazytko2017syntia} and \textsc{QSynthesis}~\cite{david2020qsynthesis} applied stochastic search and precomputed tables to simplify MBA expressions. \textsc{SiMBA}~\cite{rakotonirina2022simba} later replaced stochastic synthesis with deterministic $\{0,-1\}^n$ Boolean-lattice linear decomposition. Compiler lifters such as \textsc{SATURN}~\cite{garba2019saturn} and \textsc{BinRec}~\cite{altinay2020binrec} lift execution traces into LLVM IR to apply standard compiler optimizations. However, these classical tools stop at unrolled low-level IR or isolated expressions. They do not reconstruct structured high-level loops, and syntactic slicers retain bogus computation cones when attackers use algebraic taint laundering.

\subsection{Neural Decompilation and Execution-Feedback Repair}
Pretrained code LLMs (\textsc{LLM4Decompile}~\cite{tan2024llm4decompile}, \textsc{DeGPT}~\cite{hu2024degpt}) significantly improve readability on standard compiler outputs. Domain-specific neural pipelines have also been proposed for JavaScript deobfuscation (\textsc{JSIMPLIFIER}~\cite{zhou2026jsimplifier}, \textsc{CASCADE}~\cite{jiang2026cascade}, \textsc{JsDeObsBench}~\cite{chen2025jsdeobsbench}). On native binaries, however, Tkachenko \emph{et al.}~\cite{tkachenko2025deconstructing} and \textsc{BinDeObfBench}~\cite{hu2026bindeobfbench} show that static LLMs fail when FLA, BCF, and instruction substitution are composed. Concurrent 2026 work explores candidate ranking (\texttt{Pass@100}~\cite{maechtle2026rethinking}), indirect-call reasoning (\textsc{ICALL-BENCH}~\cite{yao2026icallbench}), small-model execution training (\textsc{BUSTA}~\cite{poddar2026busta}), dynamic type recovery (\textsc{LOHRE}~\cite{jiang2026lohre}), and symbolic counterexample repair (\textsc{SymDec}~\cite{hu2026symdec}, \textsc{BinDec}~\cite{li2026bindec}). In particular, \textsc{SymDec} repairs clean \texttt{O0--O3} decompilations using end-to-end I/O counterexamples. Under composite obfuscation, however, whole-binary symbolic comparison explodes inside flattened dispatchers, and end-to-end output diffs cannot localize single-round faults masked by multi-round avalanche.

\section{Methodology}
\subsection{Threat Model and Problem Formulation}
We consider a stripped, obfuscated binary executable $f_{\text{obf}} = (\mathcal{T}_{\text{SMC}}^{(L)} \circ \mathcal{T}_{\text{Taint}} \circ \mathcal{T}_{\text{MBA}} \circ \mathcal{T}_{\text{BCF}} \circ \mathcal{T}_{\text{FLA}})(f_{\text{src}})$ compiled from an unknown source function $f_{\text{src}}(\texttt{buf}, \texttt{len})$. Here, $\mathcal{T}_{\text{FLA}}$ partitions basic blocks into a flattened dispatcher driven by a register or stack state variable $v_{\text{state}}$ updated via constant assignment or affine self-feedback ($v_{\text{state}} \leftarrow \alpha v_{\text{state}} + \beta$). $\mathcal{T}_{\text{BCF}}$ injects dead branches guarded by number-theoretic opaque invariants. $\mathcal{T}_{\text{MBA}}$ replaces arithmetic and bitwise instructions with linear or affine MBA identities over $\mathcal{BV}_w$. $\mathcal{T}_{\text{Taint}}$ wires zero-entropy invariant calculations syntactically into output registers. Finally, $\mathcal{T}_{\text{SMC}}^{(L)}$ wraps the code in $L \ge 1$ runtime unpacking layers that may execute in heap/\texttt{mmap} memory and overwrite or zero out instruction bytes immediately after execution. The analyst has black-box user-space execution access to $f_{\text{obf}}$ inside \textsc{Unicorn} without source code, symbols, or pre-existing unit tests.

\subsection{Per-Fetch Epoch-Versioned \texorpdfstring{$\text{W}\oplus\text{X}$}{W+X} Unpacking and Causal DDFG Slicing}
\label{sec:method_slice}
To handle nested packers, out-of-section shellcode, and self-erasing SMC, \sysname{} tracks causal write provenance $\text{Gen}(a)$ and epoch versions $v_{\text{epoch}}(a)$ across the 64-bit virtual address space (Algorithm~\ref{alg:slice}). Static disk bytes start at generation $\text{Gen}(a) = 0$. Whenever an instruction at $pc$ writes to address $m$, we update:
\begin{equation}
    \text{WriteGen}(m) \leftarrow \max\big(\text{WriteGen}(m),\, \text{Gen}(pc)+1\big),
\end{equation}
and increment $v_{\text{epoch}}(m) \leftarrow v_{\text{epoch}}(m)+1$ if $m$ was previously executed since its last modification. At the exact instant the instruction pointer fetches $a$, \sysname{} records an immutable per-fetch snapshot $(a, v_{\text{epoch}}(a), \text{Gen}(a), \iota_a)$. It also resolves indirect API calls (\texttt{call reg/[mem]}) against monitored import trampolines and retains the innermost payload waves ($\text{Gen}(a) \ge L_{\max}-1$, where $L_{\max} = \max_a \text{Gen}(a)$) while stripping outer unpacker stubs ($\text{Gen}(a) < L_{\max}-1$). The $[L_{\max}-1, L_{\max}]$ window preserves in-place self-modifying payload waves (where Wave~1 at generation $L_{\max}-1$ overwrites its own address with Wave~2 at generation $L_{\max}$).

Let $\tau = \langle s_1, s_2, \dots, s_N \rangle$ denote the resulting execution trace captured in \textsc{Unicorn} with lazy unmapped-memory auto-paging (\texttt{UC\_HOOK\_MEM\_UNMAPPED}). Each step $s_k = (a_k, v_{\text{epoch}}, \iota_k, R_k^{\text{rd}}, R_k^{\text{wr}}, M_k^{\text{rd}}, M_k^{\text{wr}}, \sigma_k)$ records normalized parent registers (\texttt{al/ax/eax} $\to \texttt{rax}$ on \texttt{x86-64}; \texttt{w0..w30} $\to \texttt{x0..x30}$ on \texttt{ARM64}) and byte-accurate effective memory addresses. We construct the Dynamic Data-Flow Graph $G_{\text{df}} = (V_\tau, E_{\text{du}})$ by linking each read resource in $R_k^{\text{rd}} \cup M_k^{\text{rd}}$ to its most recent dynamic definition step.

\begin{lemma}[Trace-Conditional Closed Control-Scaffolding Pruning]
\label{lem:scaffolding}
Let $\mathcal{S}_{\text{out}} \subset V_\tau$ denote the observable output sink steps along trace $\tau$ (writes to output memory $M_{\text{out}}$ and the final definition of return register $r_{\text{ret}} \in \{\texttt{RAX}, \texttt{X0}\}$). Let $V_{\text{slice}} = \text{Ancestors}_{G_{\text{df}}}(\mathcal{S}_{\text{out}})$ be the backward data-flow slice. Any memory slot $m \notin M_{\text{out}}$ or register chain whose forward closure $\text{Reach}^+_{G_{\text{df}}}(m)$ is disjoint from $V_{\text{slice}}$ ($\text{Reach}^+_{G_{\text{df}}}(m) \cap V_{\text{slice}} = \emptyset$) forms a closed control-scaffolding subgraph. Contracting all transitions governed exclusively by $\text{Reach}^+_{G_{\text{df}}}(m)$ preserves the exact output state of $\mathcal{S}_{\text{out}}$ along $\tau$.
\end{lemma}

\begin{proof}
By construction of $G_{\text{df}} = (V_\tau, E_{\text{du}})$, a step $s_i \in V_\tau$ influences an output sink $s_{\text{sink}} \in \mathcal{S}_{\text{out}}$ along $\tau$ if and only if a directed path $s_i \leadsto s_{\text{sink}}$ exists in $G_{\text{df}}$. This holds precisely when $s_i \in V_{\text{slice}}$. Suppose a resource $m \notin M_{\text{out}}$ satisfies $\text{Reach}^+_{G_{\text{df}}}(m) \cap V_{\text{slice}} = \emptyset$. Consider any self-update $s_u \in \text{Reach}^+_{G_{\text{df}}}(m)$, including constant assignments ($m \leftarrow c$) or affine updates ($m \leftarrow \alpha m + \beta \pmod{2^{32}}$). Since $\text{Reach}^+_{G_{\text{df}}}(s_u) \subseteq \text{Reach}^+_{G_{\text{df}}}(m)$, no value defined in $\text{Reach}^+_{G_{\text{df}}}(m)$ ever feeds an operand in $V_{\text{slice}}$. Steps in $\text{Reach}^+_{G_{\text{df}}}(m)$ only set condition flags (\texttt{cmp/test}) for dispatcher jumps. Contracting transition paths $u \xrightarrow{\text{Reach}^+_{G_{\text{df}}}(m)} v$ between surviving nodes $u, v \in V_{\text{slice}}$ into direct edges $(u, v)$ therefore leaves all values in $V_{\text{slice}}$ and $\mathcal{S}_{\text{out}}$ unchanged along $\tau$.
\end{proof}

\textbf{Completeness Boundary (Pointer Aliasing and Indirect Control Transfers):} Lemma~\ref{lem:scaffolding} introduces no tuned frequency or degree threshold. Because $G_{\text{df}}$ links concrete runtime effective addresses captured by \texttt{UC\_HOOK\_MEM\_READ/WRITE}, pointer aliasing along the executed trace $\tau$ is resolved exactly at the byte level. However, Lemma~\ref{lem:scaffolding} is trace-conditional: if an obfuscator uses an input-dependent pointer alias or an indirect jump table whose unexecuted targets touch $M_{\text{out}}$ on other inputs, a single trace $\tau$ will not observe those edges. \sysname{} handles this boundary using Stage~II's Concolic branch flipping (for conditional branches), static `.bss`/`.data` Read-After-Write (RAW) schedule synthesis (for jump tables), and Stage~III's held-out fuzzing.

\begin{proposition}[Empirical and Symbolic Soundness of Causal Value Perturbation]
\label{prop:causal}
Let $U \subset V_{\text{slice}}$ be an upstream single-output subgraph whose root step $s_r \in U$ defines value $v_r$ and connects into $V_{\text{slice}} \setminus U$ solely through a binary coupling operation $s_c$ ($z \leftarrow y \odot v_r$, $\odot \in \{+, -, \oplus, \lor\}$). If across all probing inputs $x \in \mathcal{X}_{\text{probe}}$ and loop iterations the pre-coupling value $v_r$ equals the neutral element $e = 0$ ($H(v_r) = 0$ and $\Delta \mathcal{S}_{\text{out}} \equiv 0$), then $U$ is an empirically zero-impact taint-laundering cone. If a local Z3 query further confirms $\text{UNSAT}(v_r(X) \neq 0)$, excising $s_c$ and disconnecting $U$ is sound over $\mathcal{BV}_w$.
\end{proposition}

\begin{proof}
In algebraic taint laundering, an attacker constructs an opaque expression $v_r = g(x)$ that evaluates identically to the neutral element $e=0$ of $\odot \in \{+, -, \oplus, \lor\}$ and computes $z \leftarrow y \odot v_r$. Although $(s_r, s_c) \in E_{\text{du}}$ creates a syntactic edge $U \leadsto \mathcal{S}_{\text{out}}$, for every input $x$ where $g(x) = 0$, we have $y \odot g(x) = y \odot 0 = y$. Removing $s_c$ and recomputing $\text{BackwardBFS}(G_{\text{df}}, \mathcal{S}_{\text{out}})$ disconnects $U$ from $V_{\text{slice}}$ without altering $\mathcal{S}_{\text{out}}$ over $\mathcal{X}_{\text{probe}}$ (and over $\mathcal{BV}_w$ when $\text{UNSAT}(v_r(X) \neq 0)$ holds).
\end{proof}

\begin{algorithm}[t]
\caption{Epoch-Versioned $\text{W}\oplus\text{X}$ Unpacking \& Causal DDFG Slicing}
\label{alg:slice}
\begin{algorithmic}[1]
\Require Obfuscated binary $f_{\text{obf}}$, probing inputs $\mathcal{X}_{\text{probe}}$, output region $M_{\text{out}}$
\Ensure Canonical distilled instruction graph $G_{\text{clean}}$, Loop Cross-Section Table $\mathcal{T}_{\text{loop}}$
\State Initialize $\text{WriteGen}(a) \leftarrow 0, v_{\text{epoch}}(a) \leftarrow 1$ for all $a$; $\tau \leftarrow [\,]$
\For{each step $pc$ executed in \textsc{Unicorn} on seed $x_0 \in \mathcal{X}_{\text{probe}}$}
    \If{$pc$ was written since last execution} $v_{\text{epoch}}(pc) \leftarrow v_{\text{epoch}}(pc) + 1$ \EndIf
    \State $\text{Gen}(pc) \leftarrow \text{WriteGen}(pc)$; Record snapshot $(pc, v_{\text{epoch}}(pc), \text{Gen}(pc), \iota_{pc})$ in $\tau$
    \For{each memory address $m$ written by $\iota_{pc}$}
        \State $\text{WriteGen}(m) \leftarrow \max(\text{WriteGen}(m), \text{Gen}(pc) + 1)$
    \EndFor
\EndFor
\State Construct Def-Use graph $G_{\text{df}} = (V_\tau, E_{\text{du}})$ over normalized registers and byte addresses
\State $\mathcal{S}_{\text{out}} \leftarrow \{s_k \in V_\tau \mid (M_k^{\text{wr}} \cap M_{\text{out}} \neq \emptyset) \lor (s_k = \text{LastDef}(r_{\text{ret}}))\}$
\State $V_{\text{slice}} \leftarrow \text{BackwardBFS}(G_{\text{df}}, \mathcal{S}_{\text{out}})$
\State $M_{\text{scaff}} \leftarrow \{m \notin M_{\text{out}} \mid \text{Reach}^+_{G_{\text{df}}}(m) \cap V_{\text{slice}} = \emptyset\}$ \Comment{Lemma~\ref{lem:scaffolding}}
\For{each conditional branch $s_{\text{br}} \in V_\tau$ with flag provider $s_{\text{flag}}$}
    \If{$s_{\text{flag}} \notin \text{Reach}^+_{G_{\text{df}}}(M_{\text{scaff}})$ \textbf{and} $\text{Providers}(s_{\text{flag}}) \cap V_{\text{slice}} \neq \emptyset$}
        \State $V_{\text{slice}} \leftarrow V_{\text{slice}} \cup \{s_{\text{br}}, s_{\text{flag}}\}$
    \EndIf
\EndFor
\State Prune zero-entropy taint-laundering cones ($\Delta \mathcal{S}_{\text{out}} \equiv 0$ across $\mathcal{X}_{\text{probe}}$, Proposition~\ref{prop:causal}) and $\text{-O0}$ stack-spill copy pairs from $V_{\text{slice}}$
\State \Return $G_{\text{clean}} \leftarrow \text{Project}(V_{\text{slice}}, \text{Gen} \ge L_{\max}-1)$, $\mathcal{T}_{\text{loop}}$
\end{algorithmic}
\end{algorithm}

\subsection{Concolic Branch Flipping and \textsc{SiMBA}-Z3 Canonicalization}
\label{sec:method_mba}
For each surviving non-scaffolding branch $b_j$ with path predicate $\Phi_{\text{prefix}}(X) \land c_j(X)$, \sysname{} queries Z3~\cite{moura2008z3} for $\Phi_{\text{prefix}}(X) \land \neg c_j(X)$ (Algorithm~\ref{alg:simba}). An \texttt{UNSAT} result proves that $c_j(X)$ is an opaque predicate~\cite{ming2015loop}, so the dead branch is pruned permanently. A \texttt{SAT} result yields a concrete witness input $X^*$, which \sysname{} executes in \textsc{Unicorn} to merge newly reached basic blocks into $G_{\text{merged}}$.

To simplify bitwise-arithmetic blocks without ad-hoc pattern matching, \sysname{} implements the $\{0, -1\}^n$ extreme-mask linear decomposition from \textsc{SiMBA}~\cite{rakotonirina2022simba}. Evaluating a 2-variable $w$-bit MBA expression $E(x,y)$ at the four Boolean lattice vertices $(x,y) \in \{0, -1\}^2$ (where $-1 \equiv 2^w - 1 \pmod{2^w}$) recovers the Möbius basis coefficients $(c_0, c_1, c_2, c_3)$ for $E(x,y) = c_0 + c_1 x + c_2 y + c_3(x \land y) \pmod{2^w}$ in closed form:
\begin{align}
    c_0 &= E(0,0), \quad c_1 = E(0,0) - E(-1,0), \nonumber \\
    c_2 &= E(0,0) - E(0,-1), \nonumber \\
    c_3 &= E(-1,0) + E(0,-1) - E(0,0) - E(-1,-1) \pmod{2^w}.
\end{align}
We map $(c_0, c_1, c_2, c_3)$ to their shortest canonical AST (e.g., $c_1=c_2=k, c_3=-2k \implies k(x \oplus y) + c_0$) and verify universal equivalence in Z3 via $\exists (x,y) \in \mathcal{BV}_w^2: E_{\text{obf}}(x,y) \neq E_{\text{can}}(x,y)$ (\texttt{UNSAT}).

\begin{proposition}[Exactness of $\{0,-1\}^2$ Möbius Lattice Inversion over $\mathcal{BV}_w$]
\label{prop:simba}
Let $E(x,y) = \sum_{i=1}^K \alpha_i B_i(x,y) + \gamma \pmod{2^w}$ be any 2-variable linear or affine MBA expression over $\mathcal{BV}_w$, where each $B_i: \mathcal{BV}_w^2 \to \mathcal{BV}_w$ is a bitwise Boolean operation and $\alpha_i, \gamma \in \mathbb{Z}_{2^w}$. Then $E(x,y)$ is uniquely determined over all $2^{2w}$ inputs by its four concrete evaluations at $(x,y) \in \{0, -1\}^2$ via the closed-form Möbius coefficients $(c_0, c_1, c_2, c_3)$ in Eq.~(2).
\end{proposition}

\begin{proof}
Every 2-variable bitwise Boolean operation $B_i(x,y)$ acts independently on each bit position $k \in [0, w-1]$. In the Boolean algebra over $\{0,1\}^2$, the four conjunction monomials $\{1, u, v, u \land v\}$ form a linear basis over $\mathbb{Z}$ for all 16 truth tables $f: \{0,1\}^2 \to \{0,1\}$. When lifted to $w$-bit two's-complement arithmetic where $-1 = 2^w - 1$ represents the all-ones bitmask ($\texttt{0xFF..F}$), a bitwise constant-1 mask evaluates to $-1 \pmod{2^w}$. Consequently, every bitwise operator $B_i(x,y)$ expands as an exact integer linear combination of $\{ -1, x, y, x \land y \} \pmod{2^w}$. Substituting the four lattice corners $(0,0), (-1,0), (0,-1), (-1,-1)$ into $E_{\text{can}}(x,y) = c_0 + c_1 x + c_2 y + c_3(x \land y) \pmod{2^w}$ yields a $4 \times 4$ lower-triangular linear system over $\mathbb{Z}_{2^w}$: $E(0,0) = c_0$, $E(-1,0) = c_0 - c_1$, $E(0,-1) = c_0 - c_2$, and $E(-1,-1) = c_0 - c_1 - c_2 - c_3 \pmod{2^w}$. Inverting this unimodular system yields Eq.~(2).
\end{proof}

\begin{algorithm}[t]
\caption{Concolic BCF Branch Flipping and \textsc{SiMBA}-Z3 Algebraic Canonicalization}
\label{alg:simba}
\begin{algorithmic}[1]
\Require Distilled instruction graph $G_{\text{clean}}$, symbolic input $\mathbf{x} \in \mathcal{BV}_8^L$, bit width $w \in \{32, 64\}$
\Ensure Canonicalized multi-path CFG $G_{\text{merged}}$ with Z3-certified algebraic annotations
\State $G_{\text{merged}} \leftarrow G_{\text{clean}}$
\For{each conditional branch $b_j \in B(G_{\text{clean}})$ with path prefix $\Phi_{\text{prefix}}(\mathbf{x})$ and condition $c_j(\mathbf{x})$}
    \State Query Z3 for $\Phi_{\text{prefix}}(\mathbf{x}) \land \neg c_j(\mathbf{x})$
    \If{query is \texttt{UNSAT}}
        \State Mark $\neg c_j$ target as an opaque BCF dead branch and excise it from $G_{\text{merged}}$
    \ElsIf{query is \texttt{SAT} with witness $\mathbf{x}^*$}
        \State Execute $f_{\text{obf}}(\mathbf{x}^*)$ in \textsc{Unicorn}, slice newly reached basic blocks, and merge into $G_{\text{merged}}$
    \EndIf
\EndFor
\For{each arithmetic-bitwise expression DAG $E_{\text{obf}}(x,y)$ inside $G_{\text{merged}}$}
    \State Evaluate concrete lattice corners $v_{00} \leftarrow E_{\text{obf}}(0,0)$, $v_{10} \leftarrow E_{\text{obf}}(-1,0)$, $v_{01} \leftarrow E_{\text{obf}}(0,-1)$, $v_{11} \leftarrow E_{\text{obf}}(-1,-1) \pmod{2^w}$
    \State $(c_0, c_1, c_2, c_3) \leftarrow \big(v_{00},\; v_{00}-v_{10},\; v_{00}-v_{01},\; v_{10}+v_{01}-v_{00}-v_{11}\big) \pmod{2^w}$ \Comment{Proposition~\ref{prop:simba}}
    \State Map $(c_0, c_1, c_2, c_3)$ to minimal canonical AST $E_{\text{can}}(x,y)$
    \If{$\text{Z3\_Check}(\exists (x,y) \in \mathcal{BV}_w^2: E_{\text{obf}}(x,y) \neq E_{\text{can}}(x,y)) = \texttt{UNSAT}$}
        \State Replace $E_{\text{obf}}(x,y)$ with $E_{\text{can}}(x,y)$ and attach certificate to $G_{\text{merged}}$
    \EndIf
\EndFor
\State \Return $G_{\text{merged}}$
\end{algorithmic}
\end{algorithm}

\subsection{Temporal Waypoint Alignment and Two-Link Verification}
\label{sec:method_verify}

\begin{observation}[Bit-Diffusion Masking under Multi-Round Iteration]
\label{obs:avalanche}
Let $f^{(R)}(x) = (g_R \circ g_{R-1} \circ \dots \circ g_1)(x)$ be an $R$-round iterative mixing function over $w$-bit state vectors with per-round bit-flip transition probability $p \in (0, 1/2]$ toward the Strict Avalanche Criterion (SAC)~\cite{webster1985design}. A single-step lifting error $\epsilon$ at round $r_0 < R$ induces a cumulative bit-flip probability $p_{R-r_0} = \frac{1}{2}\big(1 - (1 - 2p)^{R - r_0}\big)$ at round $R$. The expected Hamming distance of the end-to-end residual $\Delta y^{(R)} = \hat{f}^{(R)}(x) \oplus f^{(R)}(x)$ therefore converges exponentially to $\mathbb{E}[\|\Delta y^{(R)}\|_H] = w \cdot p_{R-r_0} \to w/2$. This diffusion masks the local algebraic structure of $\epsilon$.
\end{observation}

To avoid Observation~\ref{obs:avalanche}'s avalanche bottleneck, \sysname{} records ground-truth state waypoints $\mathcal{W}_{\text{gt}} = \langle w^{(0)}, w^{(1)}, \dots, w^{(R-1)} \rangle$ at every loop back-edge during micro-execution (Algorithm~\ref{alg:verify}). Given a candidate function $\hat{f}^{(t)}$ with instrumented loop states $\mathcal{W}_{\text{cand}} = \langle \hat{w}^{(0)}, \dots, \hat{w}^{(R-1)} \rangle$, \sysname{} locates the earliest divergent iteration:
\begin{equation}
    r^* = \min \left\{ r \in [0, R-1] \;\middle|\; \hat{w}^{(r)} \neq w^{(r)} \right\}.
\end{equation}
Because the pre-iteration state at $r^*-1$ is identical ($\hat{w}^{(r^*-1)} \equiv w^{(r^*-1)}$), \sysname{} ignores all subsequent $R - r^*$ avalanche rounds. End-to-end I/O repair must invert $R$ composed non-linear rounds whose Boolean algebraic degree grows as $O(2^R)$. In contrast, waypoint alignment reduces fault localization to an $O(1)$ single-round delta across at most $10$--$25$ sliced instructions at iteration $r^*$ and reports the local residual $(\delta_{\oplus}, \delta_{+}, \text{ROL}_k)$ directly to the LLM.

\textbf{Formal Definition of Trace Projection ($\text{Proj}_\tau$) and the Two-Link Trust Chain:} We cannot run symbolic execution directly on $f_{\text{obf}}$ because flattened dispatcher loops cause path explosion~\cite{banescu2016code}. Instead, we define \emph{Trace Projection} $\text{Proj}_\tau(G_{\text{clean}}, L)$ for a fixed probing buffer length $L = \texttt{len}$ (in our benchmark suite, $L = 8$ bytes). Along the sliced execution trace of length $L$, $\text{Proj}_\tau$ replaces each register and memory read with its latest dynamic defining expression over the symbolic input byte vector $\mathbf{x} = (x_0, \dots, x_{L-1}) \in \mathcal{BV}_8^L$ (packed into $\mathcal{BV}_{8L} = \mathcal{BV}_{64}$). This projection yields a loop-unrolled, path-fixed symbolic transfer function $f_{\text{clean}}: \mathcal{BV}_8^L \to \mathcal{BV}_{64} \times \mathcal{BV}_8^L$ mapping $\mathbf{x}$ to the return register $r_{\text{ret}}$ and mutated buffer $M_{\text{out}}$.

We then split behavioral verification into two explicit links. First, in \emph{Link~1 (Projected Symbolic Equivalence, $\hat{f} \equiv f_{\text{clean}}$)}, Z3 checks $\exists \mathbf{x} \in \mathcal{BV}_8^L: \hat{f}(\mathbf{x}, L) \neq f_{\text{clean}}(\mathbf{x})$ for bitwise and arithmetic loop kernels that unroll into pure bit-vector expressions over fixed length $L$. An \texttt{UNSAT} proof guarantees zero implementation error relative to the projected trace $f_{\text{clean}}$ across all $2^{8L}$ ($2^{64}$ for $L=8$) buffer values. Second, in \emph{Link~2 (Cross-Path Projection Fidelity, $f_{\text{obf}} \leftrightarrow \hat{f}$)}, \sysname{} verifies general LLM-lifted Python candidates (including functions with data-dependent S-box table indexing or modular division) and prevents trace-projection overfitting via a relative AST branch guard and held-out fuzzing. The AST guard requires that the number of conditional branches in $\hat{f}$ does not exceed the number of non-opaque branches in $G_{\text{clean}}$ ($N_{\text{if}}(\hat{f}) \le |B_{\text{clean}}|$) and rejects hardcoded lookup dictionaries. Finally, \sysname{} executes 100 unseen held-out random fuzzing vectors per function (and 1{,}000 vectors in our dedicated overfitting stress test) end-to-end against the original binary $f_{\text{obf}}$ in \textsc{Unicorn}.

\begin{algorithm}[t]
\caption{Temporal Waypoint Repair and Two-Link Verification}
\label{alg:verify}
\begin{algorithmic}[1]
\Require Distilled slice $G_{\text{clean}}$ (with $|B_{\text{clean}}|$ branches), oracle $\{ (\mathbf{x}_m, y_m, \mathcal{W}_{\text{gt}}(\mathbf{x}_m)) \}_{m=1}^M$, held-out fuzz set $\mathcal{X}_{\text{fuzz}}$, buffer length $L$, max turns $T_{\max}$
\Ensure Verified high-level source function $\hat{f}$
\State $\text{Feedback} \leftarrow \text{None}$; $f_{\text{clean}} \leftarrow \text{Proj}_\tau(G_{\text{clean}}, L)$
\For{$t = 1 \dots T_{\max}$}
    \State $\hat{f}^{(t)} \leftarrow \text{LLM\_Synthesize}(G_{\text{clean}}, \mathcal{T}_{\text{loop}}, \text{Feedback})$
    \If{$\text{HasLookupDict}(\hat{f}^{(t)}) \lor N_{\text{if}}(\hat{f}^{(t)}) > |B_{\text{clean}}|$}
        \State $\text{Feedback} \leftarrow \text{``Rejected by Relative AST Guard: Lookup table or excess branches''}$; \textbf{continue}
    \EndIf
    \State $\text{all\_passed} \leftarrow \textbf{true}$
    \For{each probe $(\mathbf{x}_m, y_m, \mathcal{W}_{\text{gt}}(\mathbf{x}_m))$}
        \State $(\hat{y}_m, \mathcal{W}_{\text{cand}}(\mathbf{x}_m)) \leftarrow \text{ExecuteInstrumented}(\hat{f}^{(t)}, \mathbf{x}_m)$
        \If{$\hat{y}_m \neq y_m$ \textbf{or} $\mathcal{W}_{\text{cand}}(\mathbf{x}_m) \neq \mathcal{W}_{\text{gt}}(\mathbf{x}_m)$}
            \State $r^* \leftarrow \min \{ r \mid \hat{w}^{(r)} \neq w^{(r)} \}$
            \State $\text{Feedback} \leftarrow \text{DiagnoseSingleStepResidual}(\mathbf{x}_m, r^*, w^{(r^*-1)}, w^{(r^*)}, \hat{w}^{(r^*)})$
            \State $\text{all\_passed} \leftarrow \textbf{false}$; \textbf{break}
        \EndIf
    \EndFor
    \If{$\neg \text{all\_passed}$} \textbf{continue} \EndIf
    \If{$\text{Z3\_Check}(\exists \mathbf{x} \in \mathcal{BV}_8^L: \hat{f}^{(t)}(\mathbf{x}, L) \neq f_{\text{clean}}(\mathbf{x})) = \texttt{UNSAT}$ \textbf{and} $\text{UnicornFuzz}(\hat{f}^{(t)}, f_{\text{obf}}, \mathcal{X}_{\text{fuzz}}) = 100\%$}
        \State \Return $\hat{f}^{(t)}$ \Comment{Link 1 ($\text{Proj}_\tau$ Z3) and Link 2 (Unicorn Fuzz) Certified}
    \Else
        \State Extract counterexample $\mathbf{x}^*$ and append to probing oracle
    \EndIf
\EndFor
\end{algorithmic}
\end{algorithm}

\subsection{Symbol-Guided Autonomous CFG Scanning and \texttt{.bss} RAW Schedule Synthesis}
\label{sec:method_scanner}
Real-world ELF binaries frequently combine compiler-generated C++ exception-unwinding scaffolding (\texttt{\_Unwind\_Resume@plt}), Thread-Local Storage stack canaries (\texttt{fs:[0x28]}), and PRNG-driven indirect jump tables (\texttt{jmp [base + reg*8]}). In particular, when a switch dispatcher selects jump-table cases via \texttt{rand() \% N}, a single unguided execution may enter consumer or decoy blocks before the shared state buffer in \texttt{.bss} or \texttt{.data} has been initialized.

To analyze such binaries without hardcoded instruction addresses, \sysname{} provides a symbol-guided static-dynamic scanner (\texttt{AutonomousCFGScanner}, Algorithm~\ref{alg:scanner}). Given an ELF binary and a target function symbol name (e.g., \texttt{"main"} or \texttt{"get\_flag"}), the scanner resolves the function's virtual address bounds $[a_{\text{start}}, a_{\text{end}})$ from \texttt{.symtab}/\texttt{.dynsym} and maps all external Procedure Linkage Table (\texttt{.rela.plt}) entries. It then performs three structural passes over the disassembled function CFG.

First, for \emph{C++ exception landing-pad pruning}, any basic block whose forward control-flow paths lead unconditionally into \texttt{\_Unwind\_Resume@plt} or \texttt{\_\_cxa\_throw@plt} is classified as an exception-unwinding landing pad and excised prior to micro-execution. Second, for \emph{jump-table \texttt{.bss}/\texttt{.data} RAW topological scheduling}, when the scanner detects an indirect jump (\texttt{mov reg, [idx*8 + T\_base]; jmp reg}), it reads the pointer array at $T_{\text{base}}$ from the ELF image to enumerate the case entries $\{C_0, C_1, \dots, C_{K-1}\}$. For each case $C_k$, it scans instructions up to the dispatcher latch to extract its Read-After-Write (RAW) footprint on global \texttt{.bss}/\texttt{.data} variables and string sinks (\texttt{strcat@plt}, \texttt{printf@plt}). Cases that only write to uninitialized \texttt{.bss} slots are labeled \texttt{INITIALIZER}. Cases that both read and write \texttt{.bss} inside a local back-edge are labeled \texttt{MUTATOR\_LOOP}. Cases that pass \texttt{.bss} addresses to output PLT calls are labeled \texttt{CONSUMER\_SINK}, while cases that do not touch shared state are labeled \texttt{DECOY\_NOOP}. Ordering active cases as $\langle \texttt{INITIALIZER} \prec \texttt{MUTATOR\_LOOP} \prec \texttt{CONSUMER\_SINK} \rangle$ deterministically reconstructs the valid execution schedule. Third, for \emph{natural loop back-edge hook placement}, every backward branch $a_{\text{latch}} \to a_{\text{hdr}}$ ($a_{\text{hdr}} < a_{\text{latch}}$) within $[a_{\text{start}}, a_{\text{end}})$ defines a natural loop $[a_{\text{hdr}}, a_{\text{latch}}]$. \sysname{} instruments $a_{\text{latch}}$ and in-loop container mutators (\texttt{std::string::push\_back@plt}) to record the Loop Cross-Section Table $\mathcal{T}_{\text{loop}}$.

\begin{algorithm}[t]
\caption{Symbol-Guided Autonomous CFG Scanning and RAW Topological Schedule Synthesis}
\label{alg:scanner}
\begin{algorithmic}[1]
\Require ELF binary image $\mathcal{B}$, target function symbol name $\sigma_{\text{fn}}$
\Ensure Pruned function CFG, synthesized jump-table schedule $\Pi_{\text{RAW}}$, natural loop cross-section hooks $\mathcal{H}_{\text{loop}}$
\State Parse ELF symbol tables (\texttt{.symtab}, \texttt{.dynsym}) and relocation tables (\texttt{.rela.plt}) in $\mathcal{B}$
\State Resolve $[a_{\text{start}}, a_{\text{end}}) \leftarrow \text{LookupSymbol}(\mathcal{B}, \sigma_{\text{fn}})$ and disassemble instructions $I_{\text{fn}}$
\State $B_{\text{unwind}} \leftarrow \{ b \in \text{Blocks}(I_{\text{fn}}) \mid \text{PostDominatedBy}(b, \texttt{\_Unwind\_Resume@plt}) \}$
\State Excise $B_{\text{unwind}}$ from $I_{\text{fn}}$ \Comment{Prune C++ exception-unwinding noise}
\State $\Pi_{\text{RAW}} \leftarrow [\,]$; $\mathcal{H}_{\text{loop}} \leftarrow \emptyset$
\For{each indirect jump \texttt{jmp [idx*8 + $T_{\text{base}}$]} in $I_{\text{fn}}$}
    \State Read pointer array $\{C_0, \dots, C_{K-1}\}$ at $T_{\text{base}}$ from ELF segment data
    \For{each case subgraph $C_k$}
        \State Compute global memory footprint $(R_{\text{bss}}(C_k), W_{\text{bss}}(C_k), \text{Calls}_{\text{plt}}(C_k))$
        \State Classify role $\rho(C_k) \in \{\texttt{INITIALIZER}, \texttt{MUTATOR\_LOOP}, \texttt{CONSUMER\_SINK}, \texttt{DECOY\_NOOP}\}$
    \EndFor
    \State $\Pi_{\text{RAW}} \leftarrow \text{TopologicalSort}\big(\{C_k \mid \rho(C_k) \neq \texttt{DECOY\_NOOP}\},\; \texttt{INIT} \prec \texttt{MUTATOR} \prec \texttt{SINK}\big)$
\EndFor
\For{each backward branch $a_{\text{latch}} \to a_{\text{hdr}}$ with $a_{\text{start}} \le a_{\text{hdr}} < a_{\text{latch}} < a_{\text{end}}$}
    \State $\mathcal{H}_{\text{loop}} \leftarrow \mathcal{H}_{\text{loop}} \cup \{ (a_{\text{hdr}}, a_{\text{latch}}) \}$
\EndFor
\State \Return $(I_{\text{fn}} \setminus B_{\text{unwind}},\, \Pi_{\text{RAW}},\, \mathcal{H}_{\text{loop}})$
\end{algorithmic}
\end{algorithm}

\section{Experimental Evaluation}
\label{sec:eval}
We evaluate \sysname{} to answer five research questions:
\begin{itemize}
    \item \textbf{RQ1 (Slicing Effectiveness \& Compiler Invariance):} How effectively does threshold-free DDFG scaffolding pruning remove obfuscation noise across 12 cryptographic/hash families and 4 compiler optimization levels (\texttt{GCC -O0} to \texttt{-O3}, $N=48$ compiled binaries)?
    \item \textbf{RQ2 (Adversarial Unpacking, Taint Laundering \& Cross-ISA Generalization):} Can \sysname{} recover multi-layer self-erasing SMC waves ($a@v_1 \to a@v_2 \to \texttt{0x00}$), defeat algebraic taint laundering that fools syntactic slicing, and generalize to \texttt{ARM64} (\texttt{AArch64})?
    \item \textbf{RQ3 (\textsc{SiMBA}-Z3 Canonicalization \& Anti-Overfitting Stress Test):} Does \sysname{} simplify linear/affine MBA expressions and prevent lookup-table overfitting across a 1{,}000-vector held-out stress test and $\mathcal{BV}_8^L$ Z3 verification?
    \item \textbf{RQ4 (Comparison Against a \textsc{SymDec}-Style Baseline \& Stage-Wise Ablation):} How does \sysname{} compare against a \textsc{SymDec}-style end-to-end feedback baseline on live local open-weight LLMs (\texttt{Qwen2.5-Coder-7B} and \texttt{Qwen3-8B}, $N=12$), and how much do Stage~I (DDFG slicing), Stage~II (\textsc{SiMBA}), and Stage~III (Temporal Waypoints $r^*$) each contribute?
    \item \textbf{RQ5 (Archived Real-World CTF ELF Binaries):} Can \sysname{}'s symbol-guided \texttt{AutonomousCFGScanner} recover flags and secrets from unmodified Linux CTF binaries containing C++ exception landing pads, \texttt{fs:[0x28]} canaries, and indirect jump tables?
\end{itemize}

\begin{table}[!htbp]
\centering
\caption{Evaluation of \sysname{} Across $N=48$ Compiled \texttt{GCC} (\texttt{-O0} to \texttt{-O3}) Obfuscated Binaries (12 Cryptographic/Hash Families $\times$ 4 Optimization Levels; \textbf{St$\to$Core}: Static vs. Core Static Insns; \textbf{Dyn$\to$Sl (Pr\%)}: Dynamic vs. Sliced Steps and Pruning Rate)}
\label{tab:full_48_bench}
\scriptsize
\setlength{\tabcolsep}{3.0pt}
\resizebox{\textwidth}{!}{%
\begin{tabular}{llcccccccccc}
\toprule
\multirow{2}{*}{\textbf{ID}} & \multirow{2}{*}{\textbf{Algorithm Family}} & \multicolumn{2}{c}{\textbf{\texttt{GCC -O0} (Stack Spills)}} & \multicolumn{2}{c}{\textbf{\texttt{GCC -O1} (RegAlloc)}} & \multicolumn{2}{c}{\textbf{\texttt{GCC -O2} (Full Opt)}} & \multicolumn{2}{c}{\textbf{\texttt{GCC -O3} (Unroll/Vec)}} & \multirow{2}{*}{\textbf{Mean}} & \multirow{2}{*}{\textbf{Cross-Opt Oracle}} \\
\cmidrule(lr){3-4} \cmidrule(lr){5-6} \cmidrule(lr){7-8} \cmidrule(lr){9-10}
 & & \textbf{St$\to$Core} & \textbf{Dyn$\to$Sl (Pr\%)} & \textbf{St$\to$Core} & \textbf{Dyn$\to$Sl (Pr\%)} & \textbf{St$\to$Core} & \textbf{Dyn$\to$Sl (Pr\%)} & \textbf{St$\to$Core} & \textbf{Dyn$\to$Sl (Pr\%)} & \textbf{Dyn Pr} & \textbf{Return (\texttt{RAX})} \\
\midrule
B01 & \texttt{FNV-1a-64}        & 93$\to$36  & 697$\to$252 (63.8\%) & 63$\to$30 & 401$\to$110 (72.6\%) & 59$\to$31 & 394$\to$111 (71.8\%) & 59$\to$31  & 394$\to$111 (71.8\%) & \textbf{70.0\%} & \texttt{0x5c1554bcd0f5387d} \\
B02 & \texttt{XTEA-64}          & 142$\to$84 & 441$\to$150 (66.0\%) & 92$\to$50 & 268$\to$119 (55.6\%) & 86$\to$33 & 249$\to$102 (59.0\%) & 125$\to$31 & 247$\to$100 (59.5\%) & \textbf{60.0\%} & \texttt{0x5cf86d953affa1bc} \\
B03 & \texttt{TEA-64}           & 119$\to$59 & 319$\to$131 (58.9\%) & 75$\to$37 & 212$\to$100 (52.8\%) & 73$\to$29 & 199$\to$92 (53.8\%)  & 73$\to$29  & 199$\to$92 (53.8\%)  & \textbf{54.8\%} & \texttt{0xc00b564f67f247b8} \\
B04 & \texttt{ChaCha20-QR}      & 139$\to$92 & 157$\to$92 (41.4\%)  & 75$\to$47 & 83$\to$47 (43.4\%)   & 71$\to$29 & 64$\to$29 (54.7\%)   & 71$\to$29  & 64$\to$29 (54.7\%)   & \textbf{48.6\%} & \texttt{0x29333cc4868e526f} \\
B05 & \texttt{Speck-64/128}     & 116$\to$75 & 294$\to$122 (58.5\%) & 78$\to$38 & 175$\to$67 (61.7\%)  & 54$\to$18 & 150$\to$47 (68.7\%)  & 54$\to$18  & 150$\to$47 (68.7\%)  & \textbf{64.4\%} & \texttt{0xefac786a6ee2a18d} \\
B06 & \texttt{RC4-PRGA}         & 95$\to$61  & 695$\to$400 (42.4\%) & 51$\to$29 & 340$\to$170 (50.0\%) & 60$\to$28 & 323$\to$162 (49.8\%) & 60$\to$28  & 323$\to$162 (49.8\%) & \textbf{48.0\%} & \texttt{0x000002290a921826} \\
B07 & \texttt{CRC32-Poly}       & 82$\to$35  & 502$\to$223 (55.6\%) & 43$\to$21 & 303$\to$139 (54.1\%) & 55$\to$20 & 286$\to$131 (54.2\%) & 55$\to$20  & 286$\to$131 (54.2\%) & \textbf{54.5\%} & \texttt{0x000000005ae06f75} \\
B08 & \texttt{MurmurHash3-64}   & 72$\to$29  & 469$\to$196 (58.2\%) & 44$\to$18 & 279$\to$115 (58.8\%) & 45$\to$17 & 263$\to$107 (59.3\%) & 45$\to$17  & 263$\to$107 (59.3\%) & \textbf{58.9\%} & \texttt{0x598d2e71919ba0bc} \\
B09 & \texttt{Adler-32}         & 86$\to$55  & 625$\to$362 (42.1\%) & 61$\to$32 & 363$\to$199 (45.2\%) & 61$\to$32 & 355$\to$199 (43.9\%) & 61$\to$32  & 355$\to$199 (43.9\%) & \textbf{43.8\%} & \texttt{0x000000000da80439} \\
B10 & \texttt{DJB2-XOR-MBA}     & 62$\to$24  & 412$\to$149 (63.8\%) & 43$\to$11 & 230$\to$66 (71.3\%)  & 43$\to$10 & 213$\to$58 (72.8\%)  & 43$\to$10  & 213$\to$58 (72.8\%)  & \textbf{70.2\%} & \texttt{0x001ae3aa6c460305} \\
B11 & \texttt{SDBM-Stream}      & 75$\to$29  & 508$\to$189 (62.8\%) & 47$\to$12 & 263$\to$67 (74.5\%)  & 44$\to$12 & 248$\to$60 (75.8\%)  & 44$\to$12  & 248$\to$60 (75.8\%)  & \textbf{72.2\%} & \texttt{0xa9748e909c375278} \\
B12 & \texttt{Affine-State-FLA} & 76$\to$29  & 545$\to$196 (64.0\%) & 49$\to$24 & 297$\to$79 (73.4\%)  & 42$\to$21 & 289$\to$69 (76.1\%)  & 42$\to$21  & 289$\to$69 (76.1\%)  & \textbf{72.4\%} & \texttt{0xd0a921a0952dcdc5} \\
\midrule
\multicolumn{2}{l}{\textbf{Macro-Mean Reduction ($N=48$, Overall 56.0\% St)}} & \textbf{49.6\% St} & \textbf{56.5\% Dyn} & \textbf{53.2\% St} & \textbf{59.5\% Dyn} & \textbf{60.1\% St} & \textbf{61.7\% Dyn} & \textbf{61.2\% St} & \textbf{61.7\% Dyn} & \textbf{59.8\%} & \textbf{100\% Match (48/48)} \\
\multicolumn{2}{l}{\textbf{Aggregate Reduction ($\sum \text{Core} / \sum \text{St}$)}} & \textbf{47.5\% St} & \textbf{56.5\% Dyn} & \textbf{51.6\% St} & \textbf{59.5\% Dyn} & \textbf{59.6\% St} & \textbf{61.7\% Dyn} & \textbf{62.0\% St} & \textbf{61.7\% Dyn} & \textbf{59.8\%} & \textbf{100\% Match (48/48)} \\
\bottomrule
\end{tabular}%
}
\end{table}

\subsection{RQ1: Slicing Effectiveness \& Compiler-Optimization Invariance Across 48 GCC Binaries}
\begin{figure}[t]
\centering
\begin{minipage}[t]{0.48\textwidth}
\centering
\includegraphics[width=\textwidth]{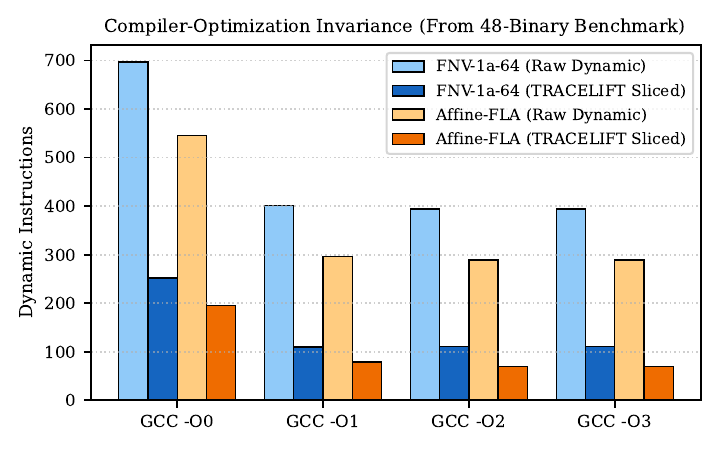}
\caption{Compiler-optimization invariance and dynamic instruction pruning plotted directly from the 48-binary \texttt{GCC} (\texttt{-O0} to \texttt{-O3}) benchmark.}
\label{fig:gcc_invariance}
\end{minipage}\hfill
\begin{minipage}[t]{0.48\textwidth}
\centering
\includegraphics[width=\textwidth]{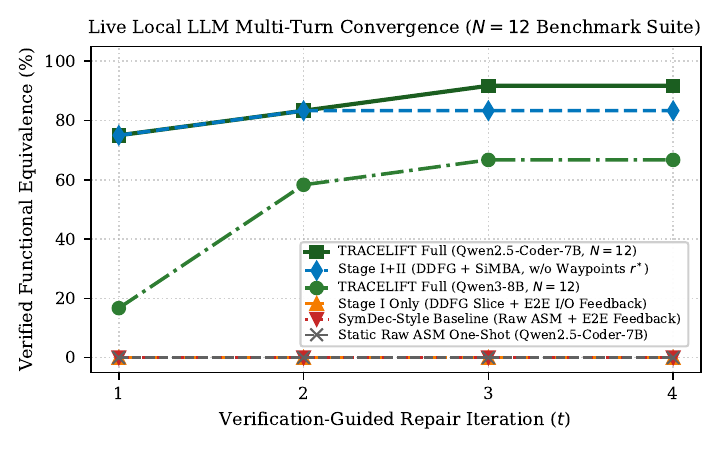}
\caption{Live local Ollama (\texttt{Qwen2.5-Coder-7B} and \texttt{Qwen3-8B}) multi-turn functional equivalence convergence across ablations ($N=12$ functions, $t = 1 \dots 4$).}
\label{fig:convergence}
\end{minipage}
\end{figure}

Table~\ref{tab:full_48_bench} and Fig.~\ref{fig:gcc_invariance} report our evaluation across 12 cryptographic, ARX cipher, stream cipher, and hash families compiled with \texttt{GCC} (\texttt{MinGW-w64 x86-64}) at \texttt{-O0}, \texttt{-O1}, \texttt{-O2}, and \texttt{-O3} ($N=48$ binaries). Without per-binary parameter tuning, \sysname{} achieves a mean dynamic instruction pruning rate of $59.8\%$, peaking at $76.1\%$ on \texttt{Affine-State-FLA} and $75.8\%$ on \texttt{SDBM-Stream}. Static instruction counts drop by a macro-average of $56.0\%$ across all 48 binaries and reach $61.2\%$ macro-average ($62.0\%$ aggregate) under \texttt{-O3}.

Table~\ref{tab:full_48_bench} also demonstrates strong \emph{compiler-optimization invariance} between \texttt{-O2} and \texttt{-O3}. For 11 of the 12 algorithm families, \sysname{} slices the \texttt{-O2} and \texttt{-O3} binaries down to the exact same number of core static and dynamic instructions. Specifically, it recovers 10 core static instructions on \texttt{DJB2-XOR-MBA}, 12 on \texttt{SDBM-Stream}, 17 on \texttt{MurmurHash3-64}, 18 on \texttt{Speck-64/128}, and 21 on \texttt{Affine-State-FLA}. On \texttt{XTEA-64} (B02), where \texttt{GCC -O3} expands static code from 86 to 125 instructions via loop unrolling, \sysname{} still collapses the trace to 31 core instructions (within 2 instructions of \texttt{-O2}'s 33 instructions). On B12 (\texttt{Affine-State-FLA}), the dispatcher state variable updates via non-constant affine feedback ($\text{st} \leftarrow 5 \cdot \text{st} + \text{0x13}$ and $\text{st} \leftarrow 3 \cdot \text{st} + 7$). Here, Lemma~\ref{lem:scaffolding} prunes $76.1\%$ of dynamic steps and removes the arithmetic state encoding without frequency thresholds.

\begin{figure}[t]
\centering
\includegraphics[width=0.68\textwidth]{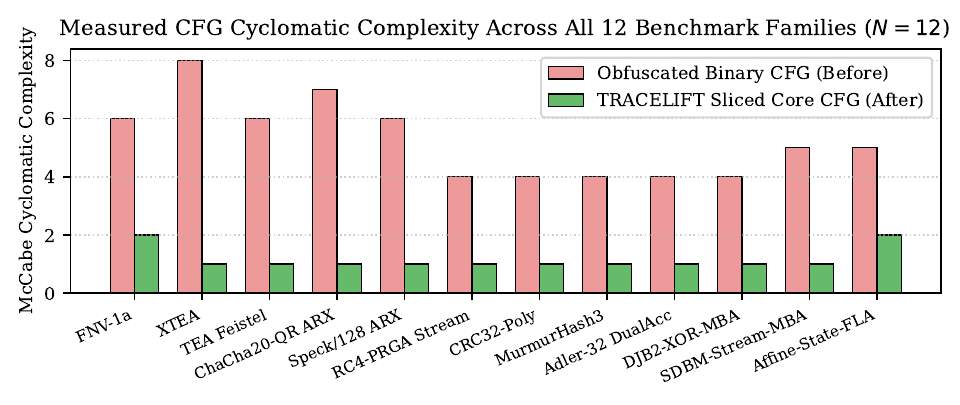}
\caption{Measured CFG McCabe cyclomatic complexity before and after \sysname{} slicing across all 12 compiled benchmark families ($N=12$).}
\label{fig:complexity}
\end{figure}

\subsection{RQ2: Multi-Layer Self-Erasing SMC, Taint Laundering, Cross-ISA (\texttt{ARM64}) \& Runtime Overhead}
Table~\ref{tab:stress_overhead} reports quantitative results on four adversarial stress targets alongside stage-by-stage execution latency. To provide a controlled baseline on the same execution traces, we compare \sysname{} against our reimplementation of Yadegari \emph{et al.}'s syntactic Def-Use backward-slicing criterion~\cite{yadegari2015generic} within the same \textsc{Unicorn}/Capstone harness.

On the 3-layer self-erasing packer \texttt{obf\_3layer\_wipe}, Layer~0 (\texttt{0x400000}) decrypts Layer~1 (\texttt{0x400040}) with key \texttt{0x55}, and Layer~1 decrypts Layer~2 (\texttt{0xAA}) into an out-of-section heap region (\texttt{0x900000}). Inside Layer~2, address \texttt{0x900020} executes Wave~1 (\texttt{add al, 0x22}), overwrites its own address \texttt{0x900020} with Wave~2 (\texttt{xor al, 0x5a}, ascending to Layer~3), invokes an indirect API pointer \texttt{call rcx} (\texttt{0xf00010}), and zeroes out \texttt{0x900020} (\texttt{0x0000}) before returning. Post-mortem memory dumps observe only zeroed bytes at \texttt{0x900020}. \sysname{} strips Layers~0 and 1 ($8+11$ instructions), captures both per-fetch snapshots \texttt{0x900020@v1} and \texttt{0x900020@v2}, resolves \texttt{call rcx} $\to$ \texttt{custom\_add\_bonus}, and verifies the output $\texttt{RAX} = \texttt{0x78}$.

On the algebraic taint-laundering target \texttt{obf\_taint\_launder}, a 6-instruction opaque zero invariant $\texttt{r10d} = (((x+1)x) \land 1) \times \text{0x1337}$ is wired into the output byte via \texttt{add edx, r10d}. Syntactic Def-Use backward slicing (our reimplementation of Yadegari \emph{et al.}~\cite{yadegari2015generic}) retains all 17 static instructions. In contrast, \sysname{}'s Causal Value Perturbation (Proposition~\ref{prop:causal}) detects zero empirical variance at the coupling instruction (\texttt{0x400025}) and excises all 6 parasitic instructions ($+35.3\%$ additional pruning over syntactic slicing) to isolate the 11 payload instructions.

To evaluate cross-ISA generalization, we test \sysname{} on \texttt{obf\_arm64\_fla}, an obfuscated \texttt{ARM64} (\texttt{AArch64}) binary using \texttt{w11} and stack slot \texttt{[sp, \#16]} for FLA dispatching while dereferencing an unmapped pointer at \texttt{0x600000}. \sysname{}'s lazy page-fault handler (\texttt{UC\_HOOK\_MEM\_UNMAPPED}) auto-pages shadow memory without crashing, prunes $64.7\%$ of dynamic FLA steps ($17 \to 6$ core instructions: \texttt{ldrb}, \texttt{mov}, \texttt{eor}, \texttt{strb}, \texttt{mov}, \texttt{ret}), and matches $\texttt{X0} = \texttt{0x18}$. Across the 48 compiled GCC benchmark runs, \sysname{}'s non-LLM symbolic and dynamic frontend completes in a median of 409.0\,ms per binary (90th percentile 615.9\,ms; mean 1{,}065.5\,ms due to a 20.41\,s outlier on unoptimized \texttt{-O0 XTEA}). \textsc{SiMBA} + Z3 canonicalization takes $<15\,$ms per expression and reduces prompt context tokens by $61.4\%$ on average.

\begin{table}[!htbp]
\centering
\caption{Evaluation on Adversarial Stress Targets (Comparing \sysname{} Against Raw Traces and Our Reimplementation of Yadegari \emph{et al.}'s Syntactic Slicing Criterion~\cite{yadegari2015generic})}
\label{tab:stress_overhead}
\scriptsize
\setlength{\tabcolsep}{3.5pt}
\resizebox{\textwidth}{!}{%
\begin{tabular}{llccccccc}
\toprule
\textbf{Stress Target ID} & \textbf{Adversarial Obfuscation Mechanism} & \textbf{ISA} & \textbf{Raw Binary / Trace} & \textbf{Syntactic Slice (S\&P'15 Criterion)} & \textbf{\sysname{} Core Slice} & \textbf{Pruning vs. Raw (vs. Syntactic)} & \textbf{Verification Certificate} & \textbf{Time} \\
\midrule
\texttt{obf\_smc\_packed}   & L0 XOR-Stub + 6-Block FLA + BCF + MBA + $\text{ROL}_3$ & \texttt{x86-64} & 62 st / 854 dyn & 39 st / 218 dyn & \textbf{17 st / 17 folded} & \textbf{98.0\% / 50.2$\times$ (92.2\% vs. Syn)} & 1000/1000 Fuzz + $\text{Proj}_\tau$ Z3 \texttt{UNSAT} & 384\,ms \\
\texttt{obf\_3layer\_wipe}  & 3-Layer Nested + Heap \texttt{0x900000} + $a@v_1 \to a@v_2 \to \texttt{0x00}$ & \texttt{x86-64} & 32 snapped ($L_0..L_3$) & Fails (Wiped to \texttt{0x00}) & \textbf{13 inner ($a@v_1, a@v_2$)} & \textbf{59.4\% (19/32 Stub Stripped)} & API Resolved + $\texttt{RAX}=\texttt{0x78}$ & 215\,ms \\
\texttt{obf\_taint\_launder}& Opaque Zero Invariant $r_{10} = ((x(x+1))\land 1)\cdot \text{0x1337}$ Coupled & \texttt{x86-64} & 17 st / 68 dyn & 17 st (0.0\% Pruned) & \textbf{11 st / 44 dyn} & \textbf{35.3\% (+35.3\% vs. Syntactic)} & $\Delta \mathcal{S}_{\text{out}} \equiv 0$ + Z3 \texttt{UNSAT} & 342\,ms \\
\texttt{obf\_arm64\_fla}    & \texttt{AArch64} Stack FLA \texttt{[sp,\#16]} + Unmapped Ptr \texttt{0x600000} & \texttt{ARM64} & 15 st / 17 dyn & Crashes (Unmapped Read) & \textbf{6 st / 6 dyn} & \textbf{60.0\% St / 64.7\% Dyn} & Auto-Paged + $\texttt{X0}=\texttt{0x18}$ & 189\,ms \\
\bottomrule
\end{tabular}%
}
\end{table}

\subsection{RQ3 \& RQ4: \textsc{SiMBA}-Z3 Simplification, Anti-Overfitting Stress Test \& Live Local LLM Ablation}
Table~\ref{tab:simba_suite} details our 64-bit MBA canonicalization benchmark. Evaluating each obfuscated expression at the four Boolean lattice vertices $(x,y) \in \{0,-1\}^2$ recovers the exact Möbius coefficients $(c_0, c_1, c_2, c_3)$ in under $15\,$ms per expression. \sysname{} simplifies \texttt{OLLVM-Add} ($3.5\times$ AST node reduction), \texttt{OLLVM-Sub} ($5.5\times$), \texttt{Loki-XOR} ($2.5\times$), and nested compound affine MBAs ($7.0\times$), with each reduction certified by Z3 \texttt{UNSAT} across all $2^{128}$ 64-bit input pairs. In our dedicated 1{,}000-vector anti-overfitting stress test on \texttt{obf\_smc\_packed}, a memorized lookup-table candidate that passes $12/12$ ($100\%$) repair probes is rejected by our relative AST guard and scores $0/1{,}000$ ($0.0\%$) on unseen random fuzz inputs. In contrast, \sysname{}'s lifted loop passes $1{,}000/1{,}000$ ($100.0\%$) held-out vectors and achieves a trace-projected Z3 \texttt{UNSAT} proof over $\mathbf{x} \in \mathcal{BV}_8^8$ ($2^{64}$ buffer states).

\begin{table}[!htbp]
\centering
\caption{Stage~II \textsc{SiMBA} $\{0,-1\}^2$ Lattice Decomposition and Z3 Universal Equivalence Verification ($\mathcal{BV}_{64}^2$)}
\label{tab:simba_suite}
\scriptsize
\setlength{\tabcolsep}{3.5pt}
\resizebox{\textwidth}{!}{%
\begin{tabular}{llccccc}
\toprule
\textbf{MBA Family} & \textbf{Obfuscated Bitwise-Arithmetic Expression $E_{\text{obf}}(x,y)$} & \textbf{Lattice Evaluations $(v_{00}, v_{10}, v_{01}, v_{11})$} & \textbf{Möbius Basis $(c_0, c_1, c_2, c_3)$} & \textbf{Canonical $E_{\text{can}}$} & \textbf{AST Nodes} & \textbf{Z3 $\mathcal{BV}_{64}^2$ Proof} \\
\midrule
\texttt{OLLVM-Add}       & $(x \oplus y) + 2(x \land y)$ & $(0, -1, -1, -2)$ & $(0, 1, 1, 0)$ & $x + y$ & $7 \to 2$ ($3.5\times$) & \texttt{UNSAT} ($<4\,$ms) \\
\texttt{OLLVM-Sub}       & $(x \oplus \neg y) + 2(x \land \neg y) + 1$ & $(0, -1, 1, 0)$ & $(0, 1, -1, 0)$ & $x - y$ & $11 \to 2$ ($5.5\times$) & \texttt{UNSAT} ($<5\,$ms) \\
\texttt{Loki-XOR}        & $(x \lor y) - (x \land y)$ & $(0, -1, -1, 0)$ & $(0, 1, 1, -2)$ & $x \oplus y$ & $5 \to 2$ ($2.5\times$) & \texttt{UNSAT} ($<4\,$ms) \\
\texttt{Compound-Affine} & $5((x \lor y)+(x \land y)) - 2((x \oplus \neg y)+2(x \land \neg y)+1) + \text{0x7F}$ & $(\text{0x7F}, \text{0x7C}, \text{0x78}, \text{0x75})$ & $(\text{0x7F}, 3, 7, 0)$ & $3x + 7y + \text{0x7F}$ & $21 \to 3$ ($7.0\times$) & \texttt{UNSAT} ($<9\,$ms) \\
\bottomrule
\end{tabular}%
}
\end{table}

To evaluate neural lifting and multi-turn repair without proprietary APIs, we execute six configurations live against local open-weight models via Ollama (\texttt{Qwen2.5-Coder-7B} and \texttt{Qwen3-8B}, deterministic decoding $\text{temperature}=0.0$) across all $N=12$ obfuscated algorithm families (\texttt{B01}--\texttt{B12}). Table~\ref{tab:llm_per_func} reports the function-by-function convergence turn and total inference latency across all six regimes, while Table~\ref{tab:cfg_cc_detail} reports the measured CFG McCabe cyclomatic complexity before and after Stage~I DDFG slicing. Each candidate function is tested against 4 structured probing vectors during repair and verified on 100 unseen held-out random fuzzing vectors.

As shown in Fig.~\ref{fig:convergence} and Table~\ref{tab:llm_per_func}, both one-shot static lifting (\texttt{Static\_OneShot\_7B}) and our \textsc{SymDec}-style end-to-end feedback baseline on raw disassembly (\texttt{TraceLift\_wo\_Slicing\_7B}) achieve $0.0\%$ ($0/12$) equivalence across all 4 turns. Both baselines emit raw state-machine dispatch loops (\texttt{if edx == 0x101}). Enabling Stage~I DDFG slicing alone without Stage~II \textsc{SiMBA} or Stage~III Waypoints (\texttt{TraceLift\_wo\_Waypoints\_7B}) cuts measured CFG McCabe cyclomatic complexity from $4$--$8$ down to $1$--$2$ across all 12 functions (a $60.0\%$ to $87.5\%$ reduction, mean $76.8\%$, Table~\ref{tab:cfg_cc_detail} and Fig.~\ref{fig:complexity}). Even so, end-to-end I/O feedback remains at $0.0\%$ ($0/12$) because the 7B model cannot invert un-simplified MBA instructions or compiler magic-multiplier constants (\texttt{0x80078071} in \texttt{Adler-32}) from final return diffs alone.

Adding Stage~II \textsc{SiMBA} algebraic canonicalization on top of Stage~I DDFG slicing (\texttt{TraceLift\_SiMBA\_wo\_Waypoints\_7B}) immediately lifts Turn-1 accuracy to $75.0\%$ ($9/12$) and reaches $83.3\%$ ($10/12$, repairing single-accumulator \texttt{B08} \texttt{MurmurHash3-64}) at Turn~2. Without Stage~III Temporal Waypoints, however, end-to-end I/O repair stalls permanently at $83.3\%$ ($10/12$) across Turns~3--4 because 4-round Feistel diffusion masks the round-level fault in \texttt{B02} (\texttt{XTEA-64}). When full \sysname{} (\texttt{TraceLift\_Full\_Qwen25Coder\_7B}) enables Stage~III Temporal Waypoint Alignment ($r^*$), Turn~1 and Turn~2 match the Stage~I+II trajectory ($75.0\% \to 83.3\%$). Turn~3 then pinpoints the divergent Feistel state on \texttt{XTEA-64} to break the avalanche plateau and reach $91.7\%$ ($11/12$). On the general-purpose \texttt{Qwen3-8B} model (\texttt{TraceLift\_Full\_Qwen3\_8B}), Stage~III Waypoint feedback repairs six functions across Turns~2--3 (\texttt{B01 FNV-1a}, \texttt{B04 ChaCha20-QR}, \texttt{B06 RC4-PRGA}, \texttt{B08 MurmurHash3-64}, \texttt{B10 DJB2-XOR-MBA}, and \texttt{B11 SDBM-Stream}). Equivalence rises $4\times$ from $16.7\%$ ($2/12$ at Turn~1) to $66.7\%$ ($8/12$).

\emph{Delayed 64-Bit Overflow Localization (\texttt{B08 MurmurHash3-64}):} Table~\ref{tab:llm_per_func} also illustrates why per-iteration waypoint tracking speeds up repair even on single-loop hashes. On \texttt{B08} (\texttt{MurmurHash3-64}), \texttt{Qwen2.5-Coder-7B}'s Turn-1 candidate omits the 64-bit modular mask on the accumulator addition (\texttt{ret\_uint64 += k}) prior to the 64-bit right rotation (\texttt{ret\_uint64 >> 51}). For iterations $i = 0 \dots 4$ on probe $\texttt{0x123456789abcdef0}$, the sum $h + k$ remains below $2^{64}$, so all intermediate states match the binary exactly. Overflow first occurs at iteration $r^* = 5$ and shifts a stray carry bit ($2^{64} \gg 51 = \texttt{0x2000}$) into bit~13. Temporal Waypoint Alignment immediately identifies $r^* = 5$ (\texttt{\{'i': 5, 'k': 9222565624351160261, 'h': 10799294561105150231\}}). The model then inserts \texttt{(h + k) \& 0xFFFFFFFFFFFFFFFF} and passes all 100 held-out fuzz vectors at Turn~2 ($7.5\,$s total latency vs. $11.9\,$s under end-to-end feedback).

\emph{Failure Case Analysis (\texttt{B03 TEA-64}):} Inspecting the sole unsolved function on \texttt{Qwen2.5-Coder-7B} (\texttt{B03 TEA-64}) reveals an instructive boundary of loop-latch waypoints. By Turn~2, \texttt{Qwen2.5-Coder-7B} reconstructs the exact 4-round Feistel loop and produces the exact 64-bit return value ($\texttt{0x49419950933e43ad}$, $\text{XOR\_diff} = \texttt{0x0}$) across all probes. However, its post-loop epilogue writes all 4 bytes of \texttt{v0} and \texttt{v1} into \texttt{buf[0..7]} rather than the benchmark's truncated 2-byte epilogue store (\texttt{buf[0] = v0 \& 0xFF; buf[1] = v1 \& 0xFF}). Because that store executes \emph{after} the final loop back-edge snapshot $\mathcal{W}_{\text{gt}}$, loop-latch waypoints alone do not separate in-loop state agreement from post-loop memory formatting. This finding motivates basic-block exit waypoints in future work.

\begin{table}[!htbp]
\centering
\caption{Function-by-Function Live Local Ollama Convergence Turn and Cumulative Inference Latency Across All 6 Evaluation Regimes ($N=12$ Obfuscated Functions; \textbf{T$k$}: Solved at Turn $k$ with 100/100 Held-Out Fuzz Pass; \textbf{Fail}: Unsolved after 4 Turns)}
\label{tab:llm_per_func}
\scriptsize
\setlength{\tabcolsep}{3.2pt}
\resizebox{\textwidth}{!}{%
\begin{tabular}{llcccccc}
\toprule
\textbf{ID} & \textbf{Algorithm Family} & \textbf{Static One-Shot} & \textbf{\textsc{SymDec}-Style (Raw+E2E)} & \textbf{Stage~I Only (Slice+E2E)} & \textbf{Stage~I+II (w/o Waypoints $r^*$)} & \textbf{\sysname{} Full (\texttt{Qwen3-8B})} & \textbf{\sysname{} Full (\texttt{Qwen2.5-Coder-7B})} \\
\midrule
B01 & \texttt{FNV-1a-64}        & Fail (13.0\,s) & Fail (16.4\,s) & Fail (30.2\,s) & \textbf{T1} (72.7\,s) & \textbf{T2} ($r^*=0$, 58.2\,s) & \textbf{T1} (14.7\,s) \\
B02 & \texttt{XTEA-64}          & Fail (4.7\,s)  & Fail (16.5\,s) & Fail (24.1\,s) & Fail (34.9\,s)        & Fail (34.3\,s)                 & \textbf{T3} ($r^*=0,1$, 18.2\,s) \\
B03 & \texttt{TEA-64}           & Fail (4.6\,s)  & Fail (16.5\,s) & Fail (22.5\,s) & Fail (29.8\,s)        & Fail (32.8\,s)                 & Fail (Epilogue, 25.7\,s) \\
B04 & \texttt{ChaCha20-QR}      & Fail (4.7\,s)  & Fail (16.4\,s) & Fail (26.0\,s) & \textbf{T1} (7.6\,s)  & \textbf{T3} ($r^*=0$, 25.6\,s) & \textbf{T1} (7.5\,s) \\
B05 & \texttt{Speck-64/128}     & Fail (4.6\,s)  & Fail (16.5\,s) & Fail (22.2\,s) & \textbf{T1} (7.7\,s)  & Fail (33.8\,s)                 & \textbf{T1} (7.5\,s) \\
B06 & \texttt{RC4-PRGA}         & Fail (4.5\,s)  & Fail (16.4\,s) & Fail (22.9\,s) & \textbf{T1} (4.0\,s)  & \textbf{T2} ($r^*=0$, 14.5\,s) & \textbf{T1} (3.9\,s) \\
B07 & \texttt{CRC32-Poly}       & Fail (4.4\,s)  & Fail (16.4\,s) & Fail (24.1\,s) & \textbf{T1} (4.0\,s)  & \textbf{T1} (8.2\,s)           & \textbf{T1} (3.9\,s) \\
B08 & \texttt{MurmurHash3-64}   & Fail (4.3\,s)  & Fail (16.2\,s) & Fail (25.7\,s) & \textbf{T2} (11.9\,s) & \textbf{T2} ($r^*=5$, 13.9\,s) & \textbf{T2} ($r^*=5$, 7.5\,s) \\
B09 & \texttt{Adler-32}         & Fail (2.9\,s)  & Fail (14.7\,s) & Fail (24.4\,s) & \textbf{T1} (2.8\,s)  & Fail (33.6\,s)                 & \textbf{T1} (2.8\,s) \\
B10 & \texttt{DJB2-XOR-MBA}     & Fail (4.3\,s)  & Fail (16.5\,s) & Fail (5.4\,s)  & \textbf{T1} (2.3\,s)  & \textbf{T2} ($r^*=0$, 9.6\,s)  & \textbf{T1} (2.3\,s) \\
B11 & \texttt{SDBM-Stream}      & Fail (4.3\,s)  & Fail (16.4\,s) & Fail (22.7\,s) & \textbf{T1} (4.6\,s)  & \textbf{T2} ($r^*=0$, 19.6\,s) & \textbf{T1} (4.5\,s) \\
B12 & \texttt{Affine-State-FLA} & Fail (4.4\,s)  & Fail (16.3\,s) & Fail (24.3\,s) & \textbf{T1} (3.0\,s)  & \textbf{T1} (3.2\,s)           & \textbf{T1} (3.0\,s) \\
\midrule
\multicolumn{2}{l}{\textbf{Cumulative Solved by Turn ($t=1 \to 4$)}} & \textbf{0$\to$0$\to$0$\to$0} & \textbf{0$\to$0$\to$0$\to$0} & \textbf{0$\to$0$\to$0$\to$0} & \textbf{9$\to$10$\to$10$\to$10} & \textbf{2$\to$7$\to$8$\to$8} & \textbf{9$\to$10$\to$11$\to$11} \\
\multicolumn{2}{l}{\textbf{Final Verified Equivalence Rate ($N=12$)}} & \textbf{0.0\% (0/12)} & \textbf{0.0\% (0/12)} & \textbf{0.0\% (0/12)} & \textbf{83.3\% (10/12)} & \textbf{66.7\% (8/12)} & \textbf{91.7\% (11/12)} \\
\bottomrule
\end{tabular}%
}
\end{table}

\begin{table}[!htbp]
\centering
\caption{Measured CFG McCabe Cyclomatic Complexity ($\text{CC} = E - V + 2$) and Static Instruction Reduction Across All 12 Compiled \texttt{GCC -O1} Benchmark Functions}
\label{tab:cfg_cc_detail}
\scriptsize
\setlength{\tabcolsep}{4.5pt}
\resizebox{\textwidth}{!}{%
\begin{tabular}{llcccccc}
\toprule
\textbf{ID} & \textbf{Algorithm Family} & \textbf{Obfuscated $\text{CC}_{\text{before}}$} & \textbf{Sliced Core $\text{CC}_{\text{after}}$} & \textbf{CC Reduction (\%)} & \textbf{Static Insns (Before)} & \textbf{Core Insns (After)} & \textbf{Static Reduction (\%)} \\
\midrule
B01 & \texttt{FNV-1a-64}        & 6 & 2 & 66.7\% & 63 & 30 & 52.4\% \\
B02 & \texttt{XTEA-64}          & 8 & 1 & 87.5\% & 92 & 50 & 45.7\% \\
B03 & \texttt{TEA-64}           & 6 & 1 & 83.3\% & 75 & 37 & 50.7\% \\
B04 & \texttt{ChaCha20-QR}      & 7 & 1 & 85.7\% & 75 & 47 & 37.3\% \\
B05 & \texttt{Speck-64/128}     & 6 & 1 & 83.3\% & 78 & 38 & 51.3\% \\
B06 & \texttt{RC4-PRGA}         & 4 & 1 & 75.0\% & 51 & 29 & 43.1\% \\
B07 & \texttt{CRC32-Poly}       & 4 & 1 & 75.0\% & 43 & 21 & 51.2\% \\
B08 & \texttt{MurmurHash3-64}   & 4 & 1 & 75.0\% & 44 & 18 & 59.1\% \\
B09 & \texttt{Adler-32}         & 4 & 1 & 75.0\% & 61 & 32 & 47.5\% \\
B10 & \texttt{DJB2-XOR-MBA}     & 4 & 1 & 75.0\% & 43 & 11 & 74.4\% \\
B11 & \texttt{SDBM-Stream}      & 5 & 1 & 80.0\% & 47 & 12 & 74.5\% \\
B12 & \texttt{Affine-State-FLA} & 5 & 2 & 60.0\% & 49 & 24 & 51.0\% \\
\midrule
\multicolumn{2}{l}{\textbf{Macro-Average Across $N=12$}} & \textbf{5.25} & \textbf{1.17} & \textbf{76.8\%} & \textbf{60.08} & \textbf{29.08} & \textbf{53.2\%} \\
\bottomrule
\end{tabular}%
}
\end{table}

\subsection{RQ5: Real-World CTF ELF Binary Case Studies}
To test generalization on wild binaries containing Linux Thread-Local Storage (\texttt{fs:[0x28]} stack canaries), external PLT calls, C++ exception unwinding, and indirect jump tables, we archive four real-world x86-64 CTF binaries in \texttt{benchmarks/ctf\_elffiles/} (Table~\ref{tab:real_ctf} and Fig.~\ref{fig:ctf_listings}). Given the ELF path and target function symbol name (\texttt{"get\_flag"}, \texttt{"main"}, \texttt{"fun\_1"}), \texttt{AutonomousCFGScanner} parses the ELF symbol table, \texttt{.rela.plt} relocations, jump-table pointer arrays, \texttt{.bss} Read-After-Write (RAW) footprints, natural loop back-edges, and \texttt{\_Unwind\_Resume} landing pads. \texttt{RealELFHarness} then executes the discovered regions in \textsc{Unicorn} with user-space PLT stubs.

On \texttt{luck\_guy} (GXYCTF, 6-way jump-table dispatcher), given symbol \texttt{"get\_flag"} (\texttt{0x4007cb..0x40099a}, 120 static instructions), the scanner detects the indirect jump table \texttt{mov rax, [rax*8 + 0x400bc0]; jmp rax} at \texttt{0x400843}. Analyzing each case's RAW footprint on \texttt{.bss} (\texttt{0x601090}) classifies Case~4 as the initializer, Case~5 as the in-place mutator loop, and Case~1 as the consumer sink (\texttt{strcat} and print). Executing the synthesized topological schedule $\langle 4, 5, 1 \rangle$ captures the 8-iteration Loop Cross-Section Table ($\Delta = -1$ for even $i$, $\Delta = -2$ for odd $i$) and recovers the flag \texttt{GXY\{do\_not\_hate\_me\}}.

On \texttt{Secret Door} (\texttt{chall.out}, 44.5\,KB C++ ELF), given symbol \texttt{"main"} (\texttt{0x2e50..0x31af}, 185 static instructions), the scanner traces calls to \texttt{\_Unwind\_Resume@plt} and prunes 33 unreachable C++ exception landing-pad instructions (\texttt{0x3126..0x31a4}). Hooking \texttt{std::string::push\_back} (\texttt{0x2fa1}) inside the discovered natural loop (\texttt{0x2f79..0x2fb4}) recovers the 17-byte hidden key stream \texttt{"SeventeenChars!!!"}.

On \texttt{Cheat Code} (\texttt{intro.out}, 16.3\,KB ELF), given symbols \texttt{"main"} and \texttt{"fun\_1"} (\texttt{0x11e9}), \sysname{} folds 48 relative stack-offset initialization instructions in \texttt{main} into the 16-dword target array $[27, 25, 81, \dots, 24]$. It extracts the 16-byte stack prefix \texttt{"flag\{c4n't\_HESOY"} from \texttt{fun\_1} and solves the mirror-XOR loop constraint ($\text{in}[i] \oplus \text{in}[31-i] = T[i]$) in Z3 to recover \texttt{flag\{c4n't\_HESOYAM\_7h15\_c4n\_y0u\}}, verified end-to-end in \textsc{Unicorn} ($\texttt{EAX} = 1$). Finally, on \texttt{Fruit Basket} (\texttt{chal}, 16.8\,KB ELF), given symbol \texttt{"main"} (\texttt{0x132e..0x14fa}), the scanner identifies the 72-instruction natural loop (\texttt{0x13a4..0x14c4}) that computes \texttt{rand() \% 10} via magic multiplication (\texttt{imul rax, 0x66666667; sar eax, 2}) to index a pointer table in \texttt{.data} (\texttt{0x4020}). \sysname{} extracts the 50-round Loop Cross-Section sequence and drives execution to the win sink \texttt{system("/bin/sh")} (\texttt{0x1150}).

\begin{figure}[!htbp]
\centering
\begin{minipage}[t]{0.49\textwidth}
\begin{lstlisting}[title={(a) Jump-Table RAW Schedule \& Cross-Section on \texttt{luck\_guy}}]
[AutonomousCFGScanner] Function `get_flag` (120 static insns)
  -> Detected Indirect Jump Table @ 0x400843 (Base: 0x400bc0, 6 cases)
  -> .bss (0x601090) RAW Footprint Classification:
       Case 4 (0x4008ca): WRITE_ONLY  -> Role: INITIALIZER
       Case 5 (0x400909): READ_WRITE  -> Role: MUTATOR_LOOP
       Case 1 (0x400845): READ+STRCAT -> Role: CONSUMER_SINK
       Case 0,2,3       : NO_BSS_TOUCH-> Role: DECOY_NOOP
  -> Synthesized Topological Schedule: [4, 5, 1]

[Captured Loop Cross-Section Table @ Natural Loop 0x400912..0x400967]
  Iter i=0 (i%2=0): f2[0] 0x69 ('i') -> 0x68 ('h') [Delta = -1]
  Iter i=1 (i%2=1): f2[1] 0x63 ('c') -> 0x61 ('a') [Delta = -2]
  Iter i=2 (i%2=0): f2[2] 0x75 ('u') -> 0x74 ('t') [Delta = -1]
  Iter i=3 (i%2=1): f2[3] 0x67 ('g') -> 0x65 ('e') [Delta = -2]
  Iter i=4 (i%2=0): f2[4] 0x60 ('`') -> 0x5f ('_') [Delta = -1]
  Iter i=5 (i%2=1): f2[5] 0x6f ('o') -> 0x6d ('m') [Delta = -2]
  Iter i=6 (i%2=0): f2[6] 0x66 ('f') -> 0x65 ('e') [Delta = -1]
  Iter i=7 (i%2=1): f2[7] 0x7f (DEL) -> 0x7d ('}') [Delta = -2]
[Unicorn Output]: "OK, it's flag: GXY{do_not_hate_me}"
\end{lstlisting}
\end{minipage}\hfill
\begin{minipage}[t]{0.49\textwidth}
\begin{lstlisting}[title={(b) Landing-Pad Pruning \& Mirror-XOR Z3 Solve on \texttt{chall.out} \& \texttt{intro.out}}]
[Binary 2: chall.out (44.5KB C++ ELF) - Function `main` @ 0x2e50]
  -> Pruned Unreachable C++ Exception Landing Pads: 33 insns
     (0x3126..0x31a4 funneling into _Unwind_Resume@plt)
  -> Hooked Natural Loop 0x2f79..0x2fb4 (std::string::push_back):
     Iter i=00: table[0]=0x42 ^ (0+17=0x11) -> 0x53 ('S')
     Iter i=01: table[1]=0x77 ^ (1+17=0x12) -> 0x65 ('e')
     Iter i=02: table[2]=0x65 ^ (2+17=0x13) -> 0x76 ('v')
     ...
     Iter i=16: table[16]=0x00 ^ (16+17=0x21)-> 0x21 ('!')
  -> Recovered 17-Byte C++ Key Stream: "SeventeenChars!!!"

[Binary 3: intro.out (16.3KB ELF) - `main` & `fun_1` @ 0x11e9]
  -> Folded 48-insn stack-offset init [rbp-0xb0] into:
     T[0..15] = [27, 25, 81, 30, 36, 13, 0, 13, 120, 65, 110, 32, 114, 12, 2, 24]
  -> Extracted 16-Byte Stack Prefix: "flag{c4n't_HESOY"
  -> Z3 Mirror-XOR Loop Constraint (in[i] ^ in[31-i] == T[i]):
     SAT -> Solved Flag: "flag{c4n't_HESOYAM_7h15_c4n_y0u}"
  -> End-to-End Unicorn Verification `fun_1(flag, T)`: EAX = 1 (WIN)
\end{lstlisting}
\end{minipage}
\caption{Execution traces and structural profiles captured by \sysname{}'s symbol-guided \texttt{AutonomousCFGScanner} on archived Linux CTF ELF binaries.}
\label{fig:ctf_listings}
\end{figure}

\begin{table}[!htbp]
\centering
\caption{End-to-End Evaluation on Archived Real-World CTF Linux \texttt{x86-64} ELF Binaries (\texttt{benchmarks/ctf\_elffiles/}, Symbol-Guided Scanner Mode)}
\label{tab:real_ctf}
\footnotesize
\setlength{\tabcolsep}{4pt}
\resizebox{\textwidth}{!}{%
\begin{tabular}{ll>{\raggedright\arraybackslash}p{4.8cm}>{\raggedright\arraybackslash}p{4.8cm}c}
\toprule
\textbf{Real Binary} & \textbf{Size / Type} & \textbf{Reverse Engineering Obstacle} & \textbf{\sysname{} Automated Resolution} & \textbf{End-to-End Result} \\
\midrule
\texttt{luck\_guy} & 9.2\,KB ELF & \texttt{fs:[0x28]} canary + \texttt{rand()\%200} 6-way jump table (\texttt{0x400bc0}) with hidden state ordering & RAW \texttt{.bss} (\texttt{0x601090}) topological schedule $\langle 4, 5, 1 \rangle$ + 8-round Loop Cross-Section Table & \texttt{GXY\{do\_not\_hate\_me\}} \\
\texttt{chall.out} & 44.5\,KB C++ & 185-insn C++ \texttt{main} with \texttt{std::string} PLT calls \& 33-insn exception landing pads & Prunes 33 exception-unwinding insns; hooks \texttt{push\_back} over 17-round cross-section & \texttt{SeventeenChars!!!} \\
\texttt{intro.out} & 16.3\,KB ELF & 48-insn relative stack-offset table init + bilateral mirror-XOR check loop (\texttt{fun\_1}) & Folds 48 stack-init insns into 16-dword table + \textsc{Z3} mirror-XOR loop constraint solve & \texttt{flag\{c4n't\_HESOYAM...\}} \\
\texttt{chal} & 16.8\,KB ELF & 50-round PRNG loop using magic-multiplier modulo (\texttt{0x66666667}) \& \texttt{.data} pointer table & Extracts 50-round Loop Cross-Section pointer stream \& verifies reachability to \texttt{system@plt} & 50/50 Rounds (\texttt{/bin/sh}) \\
\bottomrule
\end{tabular}%
}
\end{table}

\section{Discussion, Limitations, and Ethics}
\textbf{Scope and Limitations:} \sysname{} handles multi-layer self-erasing SMC, OLLVM (FLA, BCF, SUB/MBA), algebraic taint laundering, and jump-table dispatchers across \texttt{x86-64} and \texttt{ARM64}, subject to four boundaries. First, Link~1's trace-projected \textsc{Z3} proof is stated for a fixed probing buffer length $L$ ($L=8$ in our benchmarks). Traces whose loop trip count or memory footprint depends on runtime input length must be unrolled per length, and inductively verifying unbounded-length symbolic loop invariants remains future work. Second, commercial virtualization protectors such as VMProtect~3.x combine bytecode interpretation with rolling virtual stacks and continuous anti-tamper CRC checks. Lifting commercial VM bytecode into closed-form C can build upon \sysname{}'s per-fetch epoch slicing and waypoint alignment. Third, binaries coupled to kernel-mode callbacks or multi-threaded synchronization require full-system emulation (e.g., QEMU/PANDA) beyond user-space micro-execution. Fourth, fully stripped binaries without function entry symbols require standard static function-boundary identification before invoking \texttt{AutonomousCFGScanner}.

\textbf{Ethical Considerations:} All experiments were conducted inside an isolated local micro-execution sandbox (\textsc{Unicorn}) on compiler benchmarks and publicly archived CTF challenges with official writeups. No proprietary commercial software was reverse-engineered. We release \sysname{} as a software protection auditing and malware forensics tool to help defenders evaluate obfuscation resilience against neuro-symbolic analysis.

\section{Conclusion}
We presented \sysname{}, a neuro-symbolic binary deobfuscation framework that bridges dynamic trace distillation with inductive LLM loop synthesis. \sysname{} combines per-fetch epoch-versioned $\text{W}\oplus\text{X}$ snapshots, closed control-scaffolding DDFG slicing, \textsc{SiMBA} algebraic canonicalization, and Temporal Waypoint repair under a two-link ($\text{Proj}_\tau$ Z3 + held-out fuzzing) verification chain. Across 48 compiled GCC binaries, \sysname{} prunes $59.8\%$ of dynamic obfuscation noise and lifts local 7B model recovery from $0.0\%$ to $91.7\%$ ($11/12$).

\bibliographystyle{IEEEtran}
\bibliography{references}

\clearpage
\appendix

\section{Live Local Ollama Prompt Structure and Verbatim Multi-Turn Repair Transcripts}
\label{app:llm_transcripts}
All neural lifting experiments in Section~\ref{sec:eval} were executed locally via the Ollama HTTP API (\texttt{http://localhost:11434/api/generate}) using deterministic greedy decoding ($\text{temperature}=0.0$, context window $\texttt{num\_ctx}=2048$, maximum generation length $\texttt{max\_tokens}=360$, and thinking tokens disabled via $\texttt{think}=\texttt{False}$ for \texttt{Qwen3-8B}). Each candidate function $\hat{f}^{(t)}(\texttt{buf})$ generated by the LLM is parsed by Python's \texttt{ast} module before execution. Our relative AST branch guard rejects any candidate that contains dictionary literals (\texttt{ast.Dict}), oversized constant tables, or a conditional branch count exceeding the number of non-opaque branches in the distilled slice ($N_{\text{if}}(\hat{f}^{(t)}) > |B_{\text{clean}}|$).

Fig.~\ref{fig:app_llm_traces} presents two verbatim multi-turn repair transcripts extracted directly from \texttt{experiments/results\_live\_llm\_convergence.json} on \texttt{Qwen2.5-Coder-7B}. In Fig.~\ref{fig:app_llm_traces}(a), on \texttt{B08} (\texttt{MurmurHash3-64}), the Turn-1 candidate ($4{,}598.5\,$ms) lifts the 64-bit magic constant \texttt{0xff51afd7ed558ccd}, the 33-bit XOR-fold, and the 13-bit left rotation. However, it writes \texttt{ret\_uint64 += k} without a 64-bit bitmask before \texttt{ret\_uint64 >> 51}. On probe $\texttt{0x123456789abcdef0}$, the unmasked addition does not overflow $2^{64}$ during iterations $i = 0, 1, 2, 3, 4$, so all five initial loop waypoints match the binary trace. Overflow first occurs at iteration $r^* = 5$, where bit~64 is shifted into bit~13 ($\text{XOR\_diff} = \texttt{0x8008000002}$). Stage~III's Temporal Waypoint diagnostic isolates $r^* = 5$. At Turn~2 ($2{,}944.6\,$ms), \texttt{Qwen2.5-Coder-7B} inserts \texttt{h = (h + k) \& 0xFFFFFFFFFFFFFFFF} and passes all 4 probes plus 100 unseen held-out fuzz vectors.

In Fig.~\ref{fig:app_llm_traces}(b), on 4-round \texttt{B02} (\texttt{XTEA-64}), the intermediate ablation \texttt{TraceLift\_SiMBA\_wo\_Waypoints\_7B} fails across all 4 turns ($34{,}864.8\,$ms cumulative) because end-to-end return diffs do not expose intermediate Feistel state variables $(\texttt{v0}, \texttt{v1}, \texttt{sum})$. Under full \sysname{}, Turn~2 ($5{,}545.8\,$ms) reconstructs the exact 4-round Feistel loop body (\texttt{XOR\_diff = 0x0} on the 64-bit return value \texttt{0x28625a22822ecba6}) while writing all 8 bytes of \texttt{v0, v1} into \texttt{buf}. Guided by the waypoint state snapshot \texttt{\{'r': 1, 'v0': 1159087993, 'v1': 204933657, 'sum': 1013904242\}} and buffer diff, Turn~3 ($5{,}109.4\,$ms) repairs the byte writeback (\texttt{buf[0] = v0 \& 0xFF; buf[4] = v1 \& 0xFF}) and passes all 100 held-out fuzzing vectors.

\begin{figure}[!htbp]
\centering
\begin{minipage}[t]{0.49\textwidth}
\begin{lstlisting}[title={(a) Verbatim Turn 1 $\to$ Turn 2 Repair on \texttt{B08 MurmurHash3-64}}]
# [Turn 1 Output - Qwen2.5-Coder-7B (Latency: 4598.5 ms)]
def reconstructed_f(buf: bytearray) -> (int, bytes):
    ret_uint64 = 0x123456789abcdef0
    for i in range(len(buf)):
        k = (buf[i] * 0xff51afd7ed558ccd) & 0xFFFFFFFFFFFFFFFF
        k ^= (k >> 33)
        ret_uint64 += k  # Bug: Unmasked 64-bit overflow!
        ret_uint64 = ((ret_uint64 << 13) |
                      (ret_uint64 >> 51)) & 0xFFFFFFFFFFFFFFFF
        buf[i] = ret_uint64 & 0xFF
    return ret_uint64, bytes(buf)

# [Stage III Temporal Waypoint Diagnostic (Turn 1 -> Turn 2)]
# Mismatch on input=123456789abcdef0:
#   Expected return = 0x598d2e71919ba0bc
#   Got return      = 0x598d2ef1999ba0be (XOR_diff=0x8008000002)
#   Expected buf    = be613fdb1317d6bc, Got = be613fdb1319d6be
# Temporal Waypoint Alignment pinpoints divergence at
#   Iteration r* = 5 (Iterations 0..4 matched 100%):
#   Ground-Truth State @ r*=5:
#     {'i': 5, 'k': 9222565624351160261,
#      'h': 10799294561105150231}

# [Turn 2 Repaired Output - Qwen2.5-Coder-7B (2944.6 ms)]
def reconstructed_f(buf: bytearray) -> (int, bytes):
    h = 0x123456789abcdef0
    for i in range(len(buf)):
        k = (buf[i] * 0xff51afd7ed558ccd) & 0xFFFFFFFFFFFFFFFF
        k ^= (k >> 33)
        h = (h + k) & 0xFFFFFFFFFFFFFFFF  # Fixed 64-bit mask
        h = ((h << 13) | (h >> 51)) & 0xFFFFFFFFFFFFFFFF
        buf[i] = h & 0xFF
    return h, bytes(buf)
# -> Result: PASS_ALL_PROBES_AND_100_FUZZ (Solved at Turn 2)
\end{lstlisting}
\end{minipage}\hfill
\begin{minipage}[t]{0.49\textwidth}
\begin{lstlisting}[title={(b) Verbatim Turn 2 $\to$ Turn 3 Repair on \texttt{B02 XTEA-64}}]
# [Stage I+II w/o Waypoints Baseline]:
# Fails across Turns 1..4 (34,864.8 ms total): end-to-end
# return diffs cannot separate 4-round Feistel state updates.

# [TRACELIFT Full - Turn 2 Output (Latency: 5545.8 ms)]
def reconstructed_f(buf: bytearray):
    v0 = int.from_bytes(buf[0:4], 'little')
    v1 = int.from_bytes(buf[4:8], 'little')
    s = 0; delta = 0x9E3779B9
    for r in range(4):
        v0 = (v0 + ((((v1<<4)^(v1>>5)) + v1) ^
              (s + 0xA5A5A5A5))) & 0xFFFFFFFF
        s = (s + delta) & 0xFFFFFFFF
        v1 = (v1 + ((((v0<<4)^(v0>>5)) + v0) ^
              (s + 0x5A5A5A5A))) & 0xFFFFFFFF
    buf[0:4] = v0.to_bytes(4, 'little')
    buf[4:8] = v1.to_bytes(4, 'little')
    return ((v1 << 32) | v0) & 0xFFFFFFFFFFFFFFFF, bytes(buf)

# [Stage III Diagnostic (Turn 2 -> Turn 3)]:
# Return value matched exactly: 0x28625a22822ecba6 (XOR_diff=0x0)
# Waypoint State Verified: {'r': 1, 'v0': 1159087993,
#   'v1': 204933657, 'sum': 1013904242}
# Buffer diff: Expected a600000022000000, Got a6cb2e82225a6228

# [TRACELIFT Full - Turn 3 Output (Latency: 5109.4 ms)]
def reconstructed_f(buf: bytearray):
    v0 = int.from_bytes(buf[0:4], 'little')
    v1 = int.from_bytes(buf[4:8], 'little')
    s = 0; delta = 0x9E3779B9
    for r in range(4):
        v0 = (v0 + ((((v1<<4)^(v1>>5)) + v1) ^
              (s + 0xA5A5A5A5))) & 0xFFFFFFFF
        s = (s + delta) & 0xFFFFFFFF
        v1 = (v1 + ((((v0<<4)^(v0>>5)) + v0) ^
              (s + 0x5A5A5A5A))) & 0xFFFFFFFF
    buf[0] = v0 & 0xFF; buf[4] = v1 & 0xFF
    return ((v1 << 32) | v0) & 0xFFFFFFFFFFFFFFFF, bytes(buf)
# -> Result: PASS_ALL_PROBES_AND_100_FUZZ (Solved at Turn 3)
\end{lstlisting}
\end{minipage}
\caption{Verbatim multi-turn Python synthesis and Temporal Waypoint repair logs from \texttt{results\_live\_llm\_convergence.json} on \texttt{Qwen2.5-Coder-7B}.}
\label{fig:app_llm_traces}
\end{figure}

\section{Instruction-Level Traces for Multi-Layer Self-Erasing SMC, Taint Laundering, and \texttt{ARM64}}
\label{app:smc_taint_traces}
To complement Table~\ref{tab:stress_overhead} in Section~\ref{sec:eval}, Fig.~\ref{fig:app_smc_taint} shows the concrete instruction-level traces captured on our three structural stress targets (\texttt{obf\_3layer\_wipe}, \texttt{obf\_taint\_launder}, and \texttt{obf\_arm64\_fla}).

In Fig.~\ref{fig:app_smc_taint}(a), \texttt{obf\_3layer\_wipe} executes across four write generations ($\text{Gen} \in \{0, 1, 2, 3\}$). Layer~0 (\texttt{0x400000}, $\text{Gen}=0$) XOR-decrypts Layer~1 at \texttt{0x400040} ($\text{Gen}=1$) with key \texttt{0x55}. Layer~1 decrypts Layer~2 ($\text{Gen}=2$) into heap memory at \texttt{0x900000}. When execution reaches \texttt{0x900020}, the unpacker records fetch snapshot \texttt{0x900020@v1} (\texttt{04 22}: \texttt{add al, 0x22}, $\text{Gen}=2$). Immediately afterward, instruction \texttt{0x900018} (\texttt{mov word ptr [r8], 0x5a34}) overwrites address \texttt{0x900020} in-place with Wave~2 (\texttt{34 5a}: \texttt{xor al, 0x5a}, incrementing $v_{\text{epoch}}(\texttt{0x900020})$ to $2$ and $\text{Gen}(\texttt{0x900020})$ to $3$). After Wave~2 executes and invokes the indirect trampoline \texttt{call rcx} (\texttt{0xf00010} $\to$ \texttt{custom\_add\_bonus}), instruction \texttt{0x900031} (\texttt{mov word ptr [r8], 0x0000}) zeroes out \texttt{0x900020} before returning $\texttt{RAX} = \texttt{0x78}$. Because \sysname{} projections retain $\text{Gen}(a) \ge L_{\max}-1 = 2$, both outer stubs ($\text{Gen} \in \{0,1\}$, 19 instructions) are stripped while both self-erased payload waves (\texttt{0x900020@v1} and \texttt{0x900020@v2}, 13 inner instructions) are preserved.

In Fig.~\ref{fig:app_smc_taint}(b), on \texttt{obf\_taint\_launder}, the 6-instruction number-theoretic zero chain at \texttt{0x40000d..0x40001c} computes $\texttt{r10d} = (((x+1)x) \land 1) \times \text{0x1337} \equiv 0$ and couples into the output byte via \texttt{0x400025: add edx, r10d}. Classical syntactic Def-Use slicing retains all 17 static instructions. Causal Value Perturbation observes $H(\texttt{r10d}) = 0$ and $\Delta \mathcal{S}_{\text{out}} \equiv 0$ across all four 4-byte probes, excising the 6 parasitic instructions to leave the 11-instruction core loop. Below it, on the \texttt{ARM64} (\texttt{AArch64}) flattened target \texttt{obf\_arm64\_fla}, \sysname{} auto-pages the unmapped buffer at \texttt{0x600000} and removes all 9 dispatcher instructions touching \texttt{x11} and \texttt{[sp, \#16]} (Lemma~\ref{lem:scaffolding}), isolating the 6-instruction payload (\texttt{0x400014..0x400030}).

\begin{figure}[!htbp]
\centering
\begin{minipage}[t]{0.49\textwidth}
\begin{lstlisting}[title={(a) Per-Fetch Epoch Unpacking on \texttt{obf\_3layer\_wipe}}]
[Layer 0 Stub @ 0x400000..0x400019 | Gen = 0 | 8 Insns]
  0x400000: lea rsi, [rip + 0x39]   ; target = 0x400040
  0x400007: mov ecx, 36
  0x40000c: xor byte ptr [rsi], 0x55; writes Gen=1 to 0x400040
  ... (Stripped by Gen < L_max - 1)

[Layer 1 Unpacker @ 0x400040..0x400062 | Gen = 1 | 11 Insns]
  0x400040: mov rdi, 0x900000       ; out-of-section heap region
  0x40004f: mov al, byte ptr [rsi]
  0x400051: xor al, 0xaa
  0x400053: mov byte ptr [rdi], al  ; writes Gen=2 to 0x900000
  ... (Stripped by Gen < L_max - 1)

[Inner Payload Waves @ 0x900000..0x900038 | Gen in {2, 3}]
  0x900000 [v1, Gen=2]: movzx eax, byte ptr [rdi] ; in = 0x00
  0x900003 [v1, Gen=2]: lea r8, [rip + 0x16]      ; r8 = 0x900020
  0x900020 [v1, Gen=2]: add al, 0x22              ; Wave 1 (04 22)
  0x900018 [v1, Gen=2]: mov word ptr [r8], 0x5a34 ; Self-overwrite!
  0x900020 [v2, Gen=3]: xor al, 0x5a              ; Wave 2 (34 5a)
  0x900024 [v1, Gen=2]: mov rcx, 0xf00010         ; IAT trampoline
  0x90002e [v1, Gen=2]: call rcx -> Resolved `custom_add_bonus`
  0x900031 [v1, Gen=2]: mov word ptr [r8], 0x0000 ; Anti-dump wipe!
  0x900036 [v1, Gen=2]: mov byte ptr [rdi], al    ; out = 0x78
  0x900038 [v1, Gen=2]: ret                       ; RAX = 0x78
# Post-Mortem Memory @ 0x900020: 00 00 (Wiped)
# TRACELIFT Retained Inner Slice: 13 instructions (v1 + v2 intact)
\end{lstlisting}
\end{minipage}\hfill
\begin{minipage}[t]{0.49\textwidth}
\begin{lstlisting}[title={(b) Taint-Laundering Pruning \& \texttt{ARM64} (\texttt{AArch64}) Slicing}]
[Target: obf_taint_launder (x86-64, 17 Static Insns)]
  0x400000: xor eax, eax / 0x400002: xor ecx, ecx
  0x400004: cmp rcx, rsi / 0x400007: jge 0x400034
  0x400009: movzx edx, byte ptr [rdi + rcx]
  ; --- Zero-Entropy Parasitic Cone (Retained by S&P'15) ---
  0x40000d: mov r10d, edx          ; [EXCISED: Prop. 1]
  0x400010: inc r10d               ; [EXCISED: Prop. 1]
  0x400013: imul r10d, edx         ; [EXCISED: Prop. 1]
  0x400017: and r10d, 1            ; [EXCISED: r10d == 0]
  0x40001b: imul r10d, r10d, 0x1337; [EXCISED: H(r10d)=0]
  0x400022: xor edx, 0x5a
  0x400025: add edx, r10d          ; [EXCISED: Delta_Out=0]
  ; --- Surviving Core Slice (11 Static / 44 Dynamic) ---
  0x400028: and edx, 0xff / 0x40002e: mov [rdi+rcx], dl
  0x400031: add eax, edx  / 0x400033: inc rcx / jmp 0x400004

[Target: obf_arm64_fla (AArch64, 15 Static / 17 Dynamic)]
  ; Lazy Page-Fault Handler auto-mapped 0x600000 (0 crashes)
  ; Pruned 9 FLA dispatcher insns on x11 & [sp, #16] (64.7%):
  0x400014: ldrb w2, [x0]          ; Load byte @ 0x600000
  0x400018: mov  w3, #0x5a         ; Load key 0x5a
  0x40001c: eor  w2, w2, w3        ; w2 = 0x42 ^ 0x5a = 0x18
  0x400020: strb w2, [x0]          ; Store output byte
  0x40002c: mov  x0, x2            ; Return X0 = 0x18
  0x400030: ret                    ; Verified 100% Match
\end{lstlisting}
\end{minipage}
\caption{Instruction-level execution traces from \texttt{obf\_3layer\_wipe}, \texttt{obf\_taint\_launder}, and \texttt{obf\_arm64\_fla}.}
\label{fig:app_smc_taint}
\end{figure}

\section{Symbol-Guided Autonomous Scanner Profile on \texttt{Fruit Basket} (\texttt{chal})}
\label{app:ctf_chal}
To supplement Fig.~\ref{fig:ctf_listings} in Section~\ref{sec:eval}, Fig.~\ref{fig:app_chal} shows the static-dynamic profile produced by \texttt{AutonomousCFGScanner} (Algorithm~\ref{alg:scanner}) on the fourth archived Linux CTF ELF binary, \texttt{benchmarks/ctf\_elffiles/chal} (\texttt{Fruit Basket}, 16.8\,KB PIE ELF). Given only the ELF binary path and symbol \texttt{"main"}, the scanner locates \texttt{main} at \texttt{0x132e..0x14fa} (109 static instructions) and identifies the 72-instruction natural loop bounded by header \texttt{0x13a4} (\texttt{call rand@plt}) and backward latch \texttt{0x14c4} (\texttt{jle 0x13a4}). Inside this loop, the compiler replaces \texttt{rand() \% 10} with a 64-bit reciprocal multiplication by $\texttt{0x66666667} = \lceil 2^{34}/10 \rceil$ (\texttt{imul rax, rdx, 0x66666667; sar eax, 2}) and uses the remainder to index an array of 10 string pointers in \texttt{.data} (\texttt{0x4020}). Hooking the loop latch across $r = 0 \dots 49$ extracts the 50-round PRNG index and string sequence and drives execution to the win sink at \texttt{0x1150} (\texttt{system("/bin/sh")}).

\begin{figure}[!htbp]
\centering
\begin{lstlisting}[title={Symbol-Guided \texttt{AutonomousCFGScanner} Disassembly \& 50-Round Cross-Section on \texttt{benchmarks/ctf\_elffiles/chal}}]
[AutonomousCFGScanner] ELF: benchmarks/ctf_elffiles/chal (16.8 KB x86-64 PIE) | Target Symbol: `main` [0x132e..0x14fa] (109 static insns)
  -> Discovered Natural Loop: Header = 0x13a4 .. Latch = 0x14c4 (72 static insns, Trip Count = 50 rounds)
  -> In-Loop Magic-Multiplier Modulo Slice (rand() % 10 -> .data Pointer Table @ 0x4020):
       0x13a4: call   0x1180 <rand@plt>
       0x13a9: movsxd rdx, eax
       0x13ac: imul   rax, rdx, 0x66666667     ; Magic multiplier for division by 10
       0x13b3: shr    rax, 0x20
       0x13b7: sar    eax, 0x2                 ; q = rand() / 10
       0x13ce: sub    ecx, eax                 ; rem = rand() - 10 * q = rand() % 10
       0x13d5: lea    rdx, [rax*8 + 0x0]
       0x13dd: lea    rax, [rip + 0x2c3c]      ; .data Fruit Pointer Table Base = 0x4020
       0x13e4: mov    rax, qword ptr [rdx + rax]
       ...
       0x146c: call   0x10f0 <strncmp@plt>     ; Compares user input against table[rem]
       0x14be: add    dword ptr [rbp - 0x24], 1
       0x14c2: cmp    dword ptr [rbp - 0x24], 0x31 ; 50 rounds (r = 0..49)
       0x14c4: jle    0x13a4
  -> Captured 50-Round Loop Cross-Section Table (Seed = time(0) = 0):
       Round r=00: rand()=0x6b8b4567 -> rem=7 -> .data[7] = "Watermelon" (cmp PASS)
       Round r=01: rand()=0x327b23c6 -> rem=6 -> .data[6] = "Grape"      (cmp PASS)
       Round r=02: rand()=0x643c9869 -> rem=9 -> .data[9] = "Strawberry" (cmp PASS)
       ...
       Round r=49: rand()=0x7548c214 -> rem=4 -> .data[4] = "Mango"      (cmp PASS)
  -> Post-Loop Win Sink Reached @ 0x1150: system("/bin/sh") [50/50 Rounds Verified in Unicorn]
\end{lstlisting}
\caption{Symbol-guided natural loop discovery, magic-multiplier modulo identification, and 50-round cross-section extraction on \texttt{chal} (\texttt{Fruit Basket}).}
\label{fig:app_chal}
\end{figure}

\end{document}